\documentclass[prb,superscriptaddress,a4paper,twocolumn,showpacs]{revtex4}

\usepackage{amsmath,graphicx,mathrsfs}
\usepackage{mathptmx,bm,color}

\def\half{\textstyle{\frac{1}{2}}}
\def\kappaf{\kappa_{\rm f}}
\def\kB{k_{\rm B}}
\def\d{\mbox{d}}
\def\fc{f_{\rm c}}
\def\lc{l_{\rm c}}
\def\Nf{N_{\rm f}}

\begin{document}

\def\deg{$^\circ$}

\title{Stiffening of semiflexible biopolymers and cross-linked networks}
\author{ T.~van Dillen, P.~R.~Onck, and E.~Van der Giessen}
\affiliation{Micromechanics of Materials, Materials Science
Centre,University of Groningen, Nijenborgh 4, NL-9747 AG Groningen, 
The Netherlands}

%
%\date{\today}
%
\begin{abstract}

\noindent
We study the mechanical stiffening behavior in two-dimensional (2D) cross-linked 
networks of semiflexible biopolymer filaments under simple shear.  Filamental constituents 
immersed in a fluid undergo thermally excited bending motions. Pulling out these 
undulations results in an increase in the axial stiffness.  We analyze this stiffening 
behavior of 2D semiflexible filaments in detail:  we first investigate the average, 
{static} force-extension relation by considering the initially present undulated 
configuration that is pulled straight under a tensile force, and compare this result with 
the average response in which undulation dynamics is allowed during pulling, as derived 
earlier by MacKintosh and coworkers. We will show that the resulting mechanical behavior 
is rather similar, but with the axial stiffness being a factor 2 to 4 larger in the dynamic 
model. Furthermore, we study the stretching contribution in case of extensible filaments 
and show that, for 2D filaments, the mechanical response is dominated by {enthalpic} 
stretching. Based on the single-filament mechanics, we develop a 2D analytical model 
describing the mechanical behavior of biopolymer networks under simple shear, adopting 
the affine deformation assumption. These results are compared with discrete, finite-element 
(FE) calculations of a network consisting of semiflexible filaments. The FE calculations show 
that local, nonaffine filament reorientations occur that induce a transition from a bending-dominated 
response at small strains to a stretching-dominated response at larger strains. Stiffening 
in biopolymer networks thus results from a combination of stiffening in individual filaments 
and changes in the network topography.
\end{abstract}

\pacs{87.16.Ka, 87.15.La, 82.35.Lr, 82.35.Pq} \maketitle

\section{Introduction}

\noindent
The mechanics of eukaryotic cells is largely governed by the
cytoskeleton, an interpenetrated network of biopolymer protein 
filaments, spanning the region between the cell's nucleus and membrane. 
Its filamentous constituents are actin filaments, intermediate filaments 
and microtubules. It is important to gain more insight
in the physical origin of the mechanical behavior of the cytoskeleton
in view of its role in fundamental biological processes as cell
division, cell motility and mechanotransduction. It is well known that network-like biological 
tissues are compliant and respond to deformation by exhibiting an increasing stiffness,
{i.e.}, the ratio of change of stress and change of
strain. Experimental research on the viscoelastic
strain stiffening behavior in such biopolymer networks started in the
1980's, e.g. by means of micropipette and
microtwisting experiments\cite{Wang95} and rheological experiments on {in-vitro} gels
of cytoskeletal filaments (actin, vimentin, keratin),\cite{GardelScience04,Xu00,Tseng04,Janmey91,Ma99}
neuronal intermediate filaments\cite{Leterrier96} and
fibrin.\cite{Bale88,Janmey83} Through these and numerous other studies, it is
well-established by now that not only the filamental constituents but also
the type and dynamics of cross-linking proteins govern the mechanical
response of such semiflexible biopolymer networks.\cite{Wachsstock94,WagnerPNAS06,GardelPNAS06,DiDonnaPRL06,
McGrath06,Tharmann}

The mechanical behavior of semiflexible biopolymer networks has been subject 
to theoretical investigation since the mid nineties.\cite{MacKintoshPRL95,HeadPRL03,HeadPRE03,WilhelmPRL03,
Levine,DiDonnaPRE05,Storm,HeussingerPRL06}
Most studies focussed on the small-strain deformations enabling analytical treatment. The
mechanics of these networks under strain is determined by the
mechanical properties of individual filamental constituents, changes
in network topography under deformation, and finally, the molecular
properties and dynamics of cross-binding proteins. MacKintosh and
co-workers developed a model to describe the elasticity of biopolymer
networks, in which strain stiffening originates from longitudinal
stiffening in individual semiflexible filaments.\cite{MacKintoshPRL95} Filamental stiffening
is attributed to {entropic} effects, originating from the dynamic 
interactions with the surrounding (cytoplasmic) fluid.\cite{MacKintoshPRL95,KroyPRL96,Odijk,Liu}  
Next, Storm \textit{et al.} developed a continuum model by inserting this 
single-filament force-extension relation in a network of infinitely many, 
randomly oriented filaments.\cite{Storm} Under an applied shear, the network 
is then assumed to deform in an {affine} manner, allowing for an 
analytical description of the overall nonlinear network response. 

Under the affine-deformation assumption, used in network models describing
rubber elasticity,\cite{Wu93} strain stiffening of the network directly results
from stiffening in individual filaments. The validity of the affinity
assumption adopted by Storm \textit{et al.}\cite{Storm} was studied in detail by
Head \textit{et al.}\cite{HeadPRL03,HeadPRE03} and by Wilhelm and Frey\cite{WilhelmPRL03} for straight
filaments. Numerical calculations in the small-strain limit show that
network distortions are indeed affine for high density,
highly cross-linked networks, but are nonaffine for low- and intermediate-density 
networks. In a recent publication we have reported on strain stiffening in
two-dimensional (2D) cross-linked biopolymer networks comprising
discrete filaments, analyzed by the finite-element method.\cite{OnckPRL05} 
With the aid of this novel approach, the network response was calculated up to large 
strains for various filament densities, and we showed that the origin of stiffening 
lies in the network rather than in its constituents. In this discrete
filament model strain stiffening results from nonaffine network reorientations that 
induce a transition from a (relatively soft) bending-dominated response at
small strains to a (stiff) stretching dominated response at large strains. 
Next to initially straight filaments, we studied networks of initially 
undulated filaments as if instantaneous thermal undulations are frozen
in, and found that these merely postpone the onset
of the stiffening. 

Both approaches have advantages and disadvantages. The affine-network (AN) model
accounts for the dynamic interaction with the surrounding fluid during pulling,
but neglects any network rearrangements, while the discrete-network (DN) model 
accounts for the latter, but ignores the former. The aim of this article is to 
rigorously investigate the relative contribution of both effects on the overall 
stiffening by a detailed comparison of both approaches. 

This article is organized as follows. In Sect. II we will determine the undulated 
shape of 2D semiflexible filaments resulting from the interaction with the
surrounding fluid and will investigate the distribution of slack
(the relative displacement of the ends to reach a straight configuration).
Subsequently, in Sect. III we will study the mechanical behavior of
extensible and inextensible semiflexible filaments under tension. We
will in detail compare the ensemble-averaged force-extension relation
of individual, inextensible filaments with and without undulation
dynamics taken into account (to be referred to as 
the dynamic and static response, 
respectively) and show that the average mechanical
response is rather similar. In addition, we will investigate the
average static and dynamic force-extension relations in case of
extensible filaments and show that the mechanical response is largely
determined by the enthalpic stretching contribution, as confirmed by
discrete finite-element calculations. Finally, in Sect. IV we will
develop an analytical model describing the affine, mechanical behavior
of 2D biopolymer networks under simple shear. As input, the model uses
the average force-extension relation of individual filaments
determined in Sect. III. Results will be compared with the
discrete network model and we will show that nonaffine local network
reorientations, that are not taken into account in the analytical model,
play a key role in the onset of stiffening, especially at lower
densities. 
%The local network structure must therefore be taken into
%account in describing the mechanical response of biopolymer networks.

\section{Semiflexible filaments: shape and thermal undulations}

\begin{figure}[ht] % Fig. 1
\begin{center}
\includegraphics[width=0.9\columnwidth]{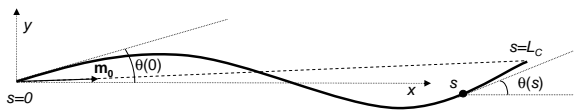}
\end{center}
\caption{Schematic of a thermally undulated filament of contour length
$L_{\rm C}$. The arc length along the filament is denoted by $s\in[0,L_{\rm C}]$ 
and its tangent relative to the $x$-axis by the angle $\theta(s)$. The unit vector along the end-to-end 
direction is $\mathbf{m}_0$.} 
\label{figfil}
\end{figure}

\noindent
The shape of cytoskeletal filaments such as F-actin is thermally
perturbed by collisions with cytoplasmic molecules. The resistance of
polymer chains to such thermal forces is characterized by their
persistence length $L_{\rm P}$, the characteristic arc length above
which the filaments' tangent becomes uncorrelated. 
Figure~\ref{figfil} shows a schematic of a snapshot
of a 2D thermally undulated filament of contour length $L_{\rm C}$,
parameterized by the arc length $s$ and having a tangent angle
$\theta(s)$.

Mathematically, the persistence length follows from the average cosine 
of $\theta(s)-\theta(0)$ in a filament over time or, alternatively, over all filaments 
in the ensemble at a specific time. For 2D filaments this average decays 
exponentially with arc length $s$ as\cite{Howard}
\begin{equation}
\left<\cos[\theta(s)-\theta(0)]\right>=\exp\left[-\frac{s}{2L_{\rm P}}\right],
\label{lp1}
\end{equation}
with the persistence length defined by
\begin{equation}
L_{\rm P}=\frac{\kappaf}{\kB T}.
\label{lp2}
\end{equation}
In Eq.~(\ref{lp2}), $k_{\rm B}$ is Boltzmann's
constant, $T$ is the temperature, and $\kappaf$ is the flexural rigidity.
While biopolymers such as DNA are flexible polymers with $L_{\rm P} \ll L_{\rm
C}$, we here focus on semiflexible filaments (e.g. actin, vimentin,
keratin and fibrin) that have a persistence length similar to their
contour length. Details on the values of $L_{\rm P}$ and $\kappaf$
used for the calculations in this article can be found in Appendix A.

In general, the undulated shape of filaments can be expressed by a superposition 
of $N$ Fourier modes for the tangent angle $\theta(s)$:~\cite{Howard,Gittes93,rmrkNOFT}
\begin{equation}
\theta(s)=\sqrt{\frac{2}{L_{\rm C}}}\sum_{n=0}^{N}a_n^0\cos\left[q_ns\right],
\label{Fourier1}
\end{equation}
with $q_n=n\pi/L_{\rm C}$. The amplitudes $a_n^0$ can be calculated
for a given shape $\theta(s)$ from
\begin{displaymath}
a_n^0=\sqrt{\frac{2}{L_{\rm C}}}\int_{0}^{L_{\rm C}}\theta(s)\cos\left[q_ns\right]\d s 
\quad\mbox{(for $n \ge 1$)}.
\end{displaymath}
Two-dimensional, undulated filaments can thus be mimicked by using Eq.~(\ref{Fourier1}) 
in which the amplitudes $a_n^0$ are randomly chosen from a Gaussian distribution with 
mean value 0 and standard deviation (see Appendix B)
\begin{equation}
s_{n}=\sqrt{\frac{1}{L_{\rm P}}}q_n^{-1} \quad (n \ge 1).
\label{standarddeviation}
\end{equation}
Since $\d x(s)=\cos\left[\theta(s)\right]\d s$ and $\d y(s)=\sin\left[\theta(s)\right]\d s$, the filament's coordinates 
$x(s)$ and $y(s)$ follow directly from
\begin{subequations}
\begin{equation}
x(s)=\int_{s'=0}^{s}\cos\left[\theta(s')\right]\d s'
\label{xparameterization}
\end{equation}
\begin{equation}
y(s)=\int_{s'=0}^{s}\sin\left[\theta(s')\right]\d s'.
\label{yparameterization}
\end{equation}
\label{parameterizations}
\end{subequations}
Finally, without loss of generality, we can require the end-to-end
direction to coincide with the $x$-axis, {i.e.}, $y(0)=y(L_{\rm
C})=0$. Application of the small-angle approximation
in Eq.~(\ref{yparameterization}), along with (\ref{Fourier1}), then yields
\begin{equation}
0 \approx \int_{0}^{L_{\rm C}}\theta(s')\d s'= \sqrt{2L_{\rm C}}a_0^0 \Rightarrow a_0^0 \approx 0.
\label{a0approx}
\end{equation}
\begin{figure}[t] %Fig. 2
\begin{center}
\includegraphics[width=0.45\textwidth]{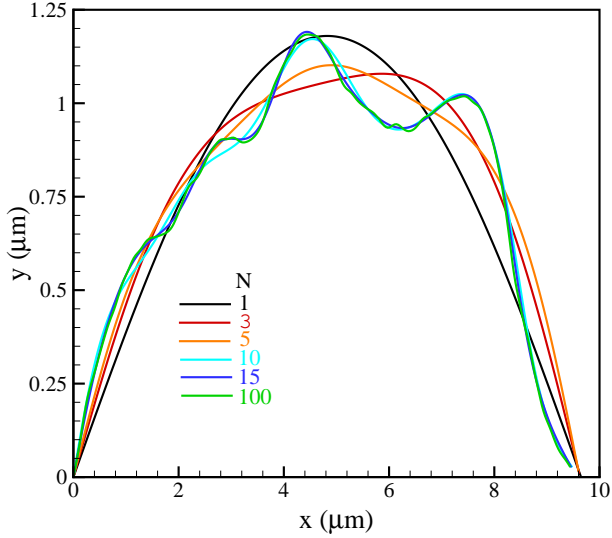}
\end{center}
\caption{A semiflexible filament, randomly generated using Eqs.~(\ref{Fourier1})--(\ref{a0approx}) 
with $L_{\rm P}=L_{\rm C}=10~\mu$m, for different numbers of
terms, $N$, in the Fourier series.}
\label{figrndfil}
\end{figure}
Figure~\ref{figrndfil} shows a filament generated with the procedure
described above using $L_{\rm P}=L_{\rm C}=10~\mu$m. The six curves all
display the filament in the same configurational state, {i.e.}, with equal (random) 
amplitudes $a_n^0$, but taking into account a varying number of terms $N$ in 
Eq. (\ref{Fourier1}). As can be seen, the shape of the filament does not significantly 
change for $N>10$.

As a result of thermal undulations, the end-to-end distance $r_{0}$ will be smaller than 
the contour length $L_{\rm C}$. To estimate the mean value of $r_{0}$ we can use the mean-squared 
end-to-end distance given by ~\cite{Howard}
\begin{eqnarray}
\left<r_{0}^{2}\right>=2\int_{s=0}^{L_{\rm C}}\int_{s'=s}^{L_{\rm C}}\exp\left[
-\frac{s'-s}{2L_{\rm P}}\right]ds'ds \nonumber\\
=4L_{\rm P}^{2}\left(2\left\{\exp\left[-\frac{L_{\rm C}}{2L_{\rm P}}\right]-1\right\}
+\frac{L_{\rm C}}{L_{\rm P}}\right)
\label{etedist}
\end{eqnarray}
\begin{figure}[t] %Fig. 3
\begin{center}
\includegraphics[width=0.45\textwidth]{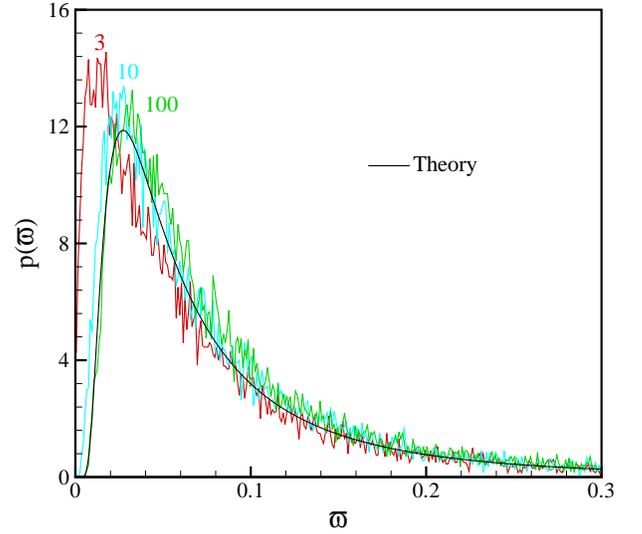}
\end{center}
\caption{Probability density function $p(\varpi)$ of the relative
slack $\varpi$ in semiflexible filaments with $L_{\rm P}=L_{\rm
C}=10~\mu$m, calculated using Eq.~(\ref{relslack}) for $2\times10^{4}$
filaments for $N=3$ (red), 10 (cyan) and 100 (green). The black, solid
curve shows the theoretical distribution function, according to Eqs.(\ref{rdf1}) and (\ref{rdf2}).}
\label{relslackdistfig}
\end{figure}
For filaments as in Fig.~\ref{figrndfil} having $L_{\rm P}=L_{\rm C}=10~\mu$m, 
$\left<r_{0}\right>$ is calculated using (\ref{etedist}) to be about $9.23~\mu$m. We have verified this by calculating 
$r_{0}=\sqrt{x^{2}(s=L_{\rm C})+y^{2}(s=L_{\rm C})}$ for $10^{5}$
filaments generated as described above with $N=10$, resulting 
in an avarage value of $r_{0}=9.25~\mu$m.\cite{remarkendpoint} 

The distance over which the ends of the filaments 
should be displaced to reach the straight configuration is $L_{\rm C}-r_0$, the {slack} 
distance $\xi$. The relative slack $\varpi$, defined as
\begin{equation}
\varpi=\frac{L_{\rm C}-r_{0}}{r_{0}},
\label{relslack}
\end{equation}
has an average value of 8.1~\% for these filaments. To investigate the dependence of $\varpi$ 
on the number of terms $N$ in the Fourier series, we plot the relative slack probability 
density function $p(\varpi)$ for  $N=3$, 10 and 100 in Fig.~\ref{relslackdistfig}. It can be seen that the distribution 
shifts to larger $\varpi$ for larger values of $N$. However, no significant difference is found 
between the distribution functions for $N=10$ and 100. 

One can also compare the relative slack distributions with the 2D
radial distribution function given by Wilhelm and Frey\cite{WilhelmPRL96},
\begin{eqnarray}
\mathscr{G}(r_0;L_{\rm C};L_{\rm P})=\frac{2L_{\rm P}\sqrt{L_{\rm
C}}}{\sqrt{\pi}}\sum_{k\ge0}\frac{(2k-1)!!}{2^kk!}
\frac{1}{[2L_{\rm P}(L_{\rm C}-r_0)]^{5/4}} \times \nonumber\\
\exp\left[-\frac{(k+\frac{1}{4})^2L_{\rm C}^2}{2L_{\rm P}(L_{\rm C}-r_0)}\right]D_{3/2}
\left(\frac{2(k+\frac{1}{4})L_{\rm C}}{\sqrt{2L_{\rm P}(L_{\rm C}-r_0)}}\right),~~	
\label{rdf1}
\end{eqnarray}
with $D_{3/2}(x)$ a parabolic cylinder function. A derivation of this
equation is given in Appendix C. This function of $r_0=L_{\rm
C}/(1+\varpi)$, normalized as
$\int_{0}^{L_{\rm C}}\mathscr{G}(r_0;L_{\rm C};L_{\rm P})\d r_0=1$, can be
transformed into the distribution function for the relative slack:
\begin{equation}
p(\varpi;L_{\rm C};L_{\rm P})=\frac{L_{\rm C}}{(1+\varpi)^2}\mathscr{G}\left(\frac{L_{\rm C}}
{1+\varpi};L_{\rm C};L_{\rm P}\right).
\label{rdf2}
\end{equation}
The function $p(\varpi;10~\mu\mbox{m};10~\mu\mbox{m})$ is plotted in Fig.~\ref{relslackdistfig} 
(solid, black curve) and agrees well with the generated distributions for large $N$.

\section{The static and dynamic mechanical behavior of semiflexible filaments}

\noindent
In this section we study the mechanical response of individual semiflexible filaments subjected 
to a tensile load. As mentioned in Sect. I, this behavior depends on whether the undulation 
dynamics is taken into account. In addition, we investigate the enthalpic contribution of 
stretching in extensible filaments.

We start by investigating the purely {static}\cite{rmrkstatic}
mechanical behavior of originally undulated, inextensible
filaments. In the absence of an externally applied force, the chain
in an undulated configuration is characterized by the set of
amplitudes \{$a_n^0$\}, as
descibed in Sect. II. Next, the chain is subjected to a  tensile force
$\fc$ along its end-to-end direction (the $x$-axis in
Fig.~\ref{figrndfil}), which induces a bending moment that tends to
flatten out the filament. As a result, the configurational state will
change to \{$a_n(\fc)$\} with a decrease in the absolute value of each
mode amplitude, {i.e.}, $|a_n(\fc\ge0)|\le |a_n^0|$. Describing
the shape of the filament at each force level $\fc$ as a Fourier series 
according to Eq.~(\ref{Fourier1}), the transverse deflection $y(s;\fc)$ is given by
\begin{equation}
y(s;\fc)=\sqrt{\frac{2}{L_{\rm C}}}\sum_{n\ge1}\frac{a_{n}(\fc)}{q_n}\sin\left[q_n s\right].
\label{deflection}
\end{equation}
The dependence of each mode amplitude on the applied force can be
calculated using the beam equation as follows. For an external force $\fc$ ,
the induced bending moment at a point $s$ along the contour is given by
$\fc y(s;\fc)$, which has to be equal to the bending moment
corresponding to the curvature of the filament, {i.e.},
\begin{equation}
\fc y(s;\fc)=\kappaf\frac{\d \theta(s;\fc)}{\d s}.
\label{beamequation}
\end{equation}
Substitution of Eq.~(\ref{Fourier1}) (using $a_n(\fc)$ instead of $a_n^0$) and Eq.~(\ref{deflection}) 
into Eq.~(\ref{beamequation}) results in
\begin{equation}
(\fc+\kappaf q_n^2)a_n=0~~~ \mbox (n \ge 1).
\label{anequation1}
\end{equation}
By increasing the external force from $\fc$ to $\fc+\d \fc$, mode amplitudes change to $a_n+\d a_n$. We can 
then apply Eq.~(\ref{anequation1}) to find a linear, first-order differential equation for each 
mode amplitude $a_n(\fc)$:
\begin{equation}
\frac{\d a_n}{a_n}=
-\frac{\d \fc}{\fc+\kappaf q_n^2}. \qquad 
\label{anequation2}
\end{equation}
With the initial condition $a_n(\fc=0)=a_n^0$, the solution reads
\begin{equation}
a_n(\fc)=\frac{\kappaf q_n^2}{\fc+\kappaf q_n^2}a_n^0.
\label{anequation3}
\end{equation}

Equations~(\ref{Fourier1}) and (\ref{anequation3}) describe the shape of the filament at a given 
applied force, which enables us to calculate the force-dependent end-to-end distance. According 
to Eq.~(\ref{xparameterization}), the end-to-end distance in the small-angle approximation is
\begin{displaymath}
r(\fc)=x(s=L_{\rm C};\fc) \approx L_{\rm C}-\frac{1}{2}\int_{0}^{L_{\rm C}}\theta^2(s)\d s,
\end{displaymath}
which, after substitution of the Fourier series for $\theta$, yields
\begin{equation}
r(\fc) = L_{\rm C}-\frac{1}{2}\sum_{n\ge1}a_n^2(\fc)
\label{fcdepeted}
\end{equation}
or, with the aid of (\ref{anequation3}),
\begin{equation}
r(\fc) = L_{\rm C}-\frac{1}{2}\sum_{n\ge1}\frac{\kappaf^2
q_n^4}{(\fc+\kappaf q_n^2)^2}(a_n^0)^2 \,.
\label{fcdepeted2}
\end{equation}
Alternatively, we can cast the force-extension relation
$u(\fc)=r(\fc)-r_0=r(\fc)-r(0)$ in the form
\begin{equation}
u(\fc)= \frac{1}{2}\sum_{n\ge1}\frac{\fc^2+2\fc\kappaf
q_n^2}{(\fc+\kappaf q_n^2)^2}(a_n^0)^2.
\label{ferstatic}
\end{equation}

\begin{figure}[t] %Fig 4
\begin{center}
\includegraphics[width=0.45\textwidth]{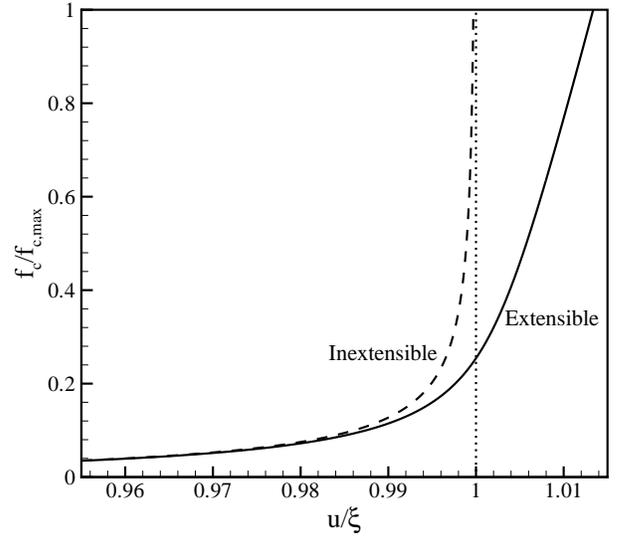}
\end{center}
\caption{The force-extension relation of a randomly generated, inextensible filament 
(dashed curve) and extensible filament (solid curve), with $L_{\rm P}=L_{\rm C}=10~\mu$m 
and $\kappaf=8.53\times10^{-17}~\mbox{Nm}^2$. The displacement is normalized with the filament's
 slack distance $\xi$ ($=0.457~\mu$m) and the force is normalized with a value 
of $f_{\rm c,max}=0.01~\mbox{N}$. The dotted line indicates the vertical asymptote for 
inextensible filaments. The value of the stretching stiffness used for the extensible 
filament is $\mu=16~\mbox{N}$.}
\label{feranalytical}
\end{figure}

Figure \ref{feranalytical} shows the mechanical response, $\fc(u)$,
according to Eq.~(\ref{ferstatic}) of a
randomly generated, inextensible filament with $L_{\rm P}=L_{\rm C}=10~
\mu$m and $\kappaf=8.53\times10^{-17}~\mbox{Nm}^2$ (dashed
curve). The displacement $u$ is normalized with the
slack distance $\xi=L_{\rm C}-r(0)=\frac{1}{2}\sum_{n\ge1}(a_n^0)^2$,
having a value of $0.457~\mu$m for this specific filament, and the
force is normalized with $f_{\rm c,max}=0.01~\mbox{N}$. 
Initially, undulations are being pulled out at relatively small
forces, but as the displacement increases and the filament becomes
(close to being) straight ({i.e.}, $u\rightarrow \xi$ or $r
\rightarrow L_{\rm C}$), the filament locks and the force diverges,
since the inextensibility allows for no additional axial straining
(see vertical, dotted line). If, however, a chain has a finite
stretching stiffness $\mu$, it can be stretched 
beyond its slack distance $\xi$. In this case, the force-dependent 
end-to-end length $r_{\mu}(\fc)$ can be constructed from Eq.~(\ref{fcdepeted2}) 
for inextensible filaments, by applying the transformation
\begin{equation}
r_{\mu}(\fc) = \left[1+\frac{\fc}{\mu}\right] \times r\left(\fc\left[1+\frac{\fc}
{\mu}\right]\right),
\label{fcdepeted3}
\end{equation}
according to Storm and coworkers.\cite{Storm} The resulting force-extension 
relation for $\mu=16~\mbox{N}$ (see Appendix A) is plotted in Fig.~\ref{feranalytical} (solid line). For small displacements the mechanical 
response of the extensible and inextensible filament is identical, since this
behavior is then dominated by internal bending. However, as the filament 
straightens, the force increases and the filament extends due to its nonzero compliance 
({i.e.}, finite stretching stiffness). For $u/\xi \ge 1.0$, the mechanical 
response of the filament gradually becomes linear with a stiffness $\d \fc/\d u$ approaching 
$\mu/L_{\rm C}$.\cite{rmrkfinalstiffness} As a check, the force-extension relation was 
also calculated numerically by dividing the filament into 500 Euler-Bernoulli beam 
elements of equal  length (see reference~ \cite{OnckPRL05} for details); the force-extension curves 
in Fig.~\ref{feranalytical} were reproduced to within 2\%. 

Now that we have determined the characteristics of the initial shape
of an individual filament and its mechanical response to an externally
applied tensile force, we can determine the {ensemble-averaged}
force-extension relation of inextensible filaments. In the static
approach adopted so far, the interaction of chains with the
surrounding fluid is taken into account only via their initial,
undulated configurational state, as can be seen from
Eq.~(\ref{fcdepeted2}). The chain's initial free energy is given by the Hamiltonian
\begin{equation}
\mathscr{H}_{\rm B}=\frac{1}{2}\kappaf\int_{0}^{L_{\rm C}}\left(\frac{\d \theta}{\d s}\right)^2\d s,
\label{hamiltonian}
\end{equation}
for the bending energy in the absence of tension ($\fc=0$). By substituting Eq.~(\ref{Fourier1}) 
into Eq.~(\ref{hamiltonian}) and by setting each harmonic energy mode equal to $\kB T/2$ 
(equipartition theorem), the mean-squared value of $a_n^0$ becomes
\begin{equation}
\left<(a_n^0)^2\right>=\frac{1}{L_{\rm P}q_n^2}.
\label{meansquaredan1}
\end{equation}
This result can be substituted into Eq.~(\ref{fcdepeted2}), yielding
the force-dependent, ensemble-averaged end-to-end distance
\begin{equation}
\left<r\right>(\varphi)=L_{\rm C}-\frac{L_{\rm C}^2}{2\pi^2L_{\rm P}}\sum_{n \ge 1}
\frac{n^2}{(n^2+\varphi)^2}, \quad
\varphi \equiv \fc L_{\rm C}^2/(\kappaf\pi^2)
\label{fcdepetedAVG1}
\end{equation}
where $\varphi$ is the force normalized with the critical Euler buckling force, i.e.,
$\varphi \equiv \fc L_{\rm C}^2/(\kappaf\pi^2)$. The summation in
Eq.~(\ref{fcdepetedAVG1}) up to $N \to \infty$ can be carried out analytically, giving
\begin{equation}
\left<r\right>(\varphi)=L_{\rm C}-\frac{L_{\rm C}^2}{8L_{\rm P}}\left(\frac{\coth[\pi 
\sqrt{\varphi}]}{\pi \sqrt{\varphi}}-\sinh ^{-2}[\pi \sqrt{\varphi}]\right).
\label{fcdepetedAVG2}
\end{equation}

\begin{figure}[t] % Fig. 5
\begin{center}
\includegraphics[width=0.45\textwidth]{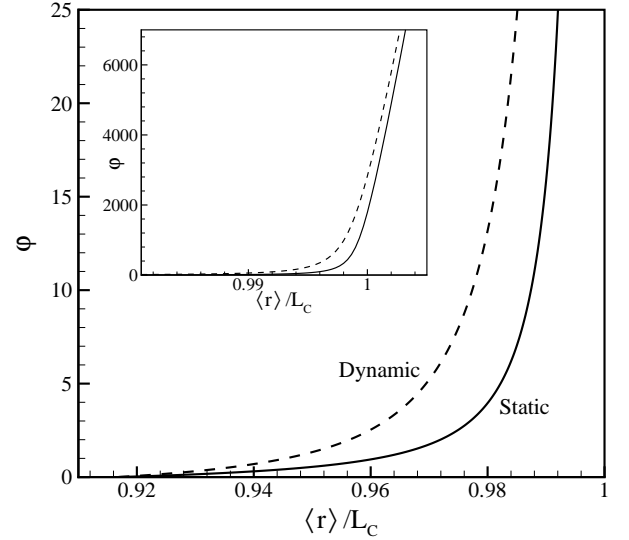}
\end{center}
\caption{The ensemble-averaged mechanical response of inextensible semiflexible
filaments having $L_{\rm P}=L_{\rm C}=10 \mu$m and
$\kappaf=8.53\times10^{-17}~\mbox{Nm}^2$ in terms of the end-to-end
distance $\left<r\right>$ versus the dimensionless force 
$\varphi=\fc L_{\rm C}^2/(\kappaf\pi^2)$. Results are plotted for the static model, using 
Eq.~(\ref{fcdepetedAVG2}) [solid curve], and for the dynamic model using Eq.~(\ref{fcdepetedAVG6}) 
[dashed curve]. The inset shows the corresponding average mechanical response of extensible 
semiflexible filaments, using Eq.~(\ref{fcdepeted3}) with $\mu=16~\mbox{N}$.}
\label{feravg}
\end{figure}

Figure \ref{feravg} shows the static, mechanical behavior according to Eq.~(\ref{fcdepetedAVG2}) [solid curve]. 
The response is linear at small forces and diverges as the average end-to-end distance 
approaches the contour length. The linear response at small forces can be found by a Taylor expansion 
of Eq.~(\ref{fcdepetedAVG2}),
\begin{equation}
\left<r\right>(\varphi) \approx \left(L_{\rm C}-\frac{L_{\rm C}^2}{12L_{\rm P}}\right) + 
\frac{L_{\rm C}^2 \pi^2 \varphi}{90L_{\rm P}}.
\label{fcdepetedAVG3}
\end{equation}
First, we conclude from this, or directly from $\left<r\right>(0)$ in
Eq.~(\ref{fcdepetedAVG1}), that the average slack distance for 2D
filaments is $\left<\xi\right>=L_{\rm C}^2/(12L_{\rm P})$. This value is also obtained by expanding
the square root of the result in Eq.~(\ref{etedist}) in a Taylor series.\cite{rmrkexpansion} For $L_{\rm C}=
L_{\rm P}=10~\mu$m the initial, average end-to-end distance is $9.17~\mu$m, close to the value of $9.25~\mu$m 
we found in Sect II. Second, the initial value of the axial stiffness 
$G_{\rm b} \equiv \d\fc/\d\left<r\right>$
resulting from internal bending by 
pulling out the thermal undulations is found to be
\begin{equation}
G_{\rm b}^0=G_{\rm b}\left(\left<r\right>=\left<r_0\right>\right)=\left(\frac{\d \left<r\right>}{\d \varphi}\frac{\d \varphi}{\d \fc}\right)^{-1}
\Biggr|_{\fc=0}=\frac{90L_{\rm P}\kappaf}{L_{\rm C}^{4}}.
\label{bendingstiffness1}
\end{equation}
The diverging behavior for large forces can be found from
Eq.~(\ref{fcdepetedAVG2}). Since $\coth[\pi
\sqrt{\varphi}] \to 1$ and $\sinh ^{-2}[\pi \sqrt{\varphi}] \to 0$ for large $\varphi$, the limiting behavior can be expressed as
\begin{equation}
\left<r\right>(\varphi) \rightarrow L_{\rm C}-\frac{L_{\rm C}^2}{8L_{\rm P}\pi \sqrt{\varphi}},
\label{fcdepetedAVG4}
\end{equation}
implying that the force-extension relation diverges as $\fc \propto
\left(L_{\rm C}-\left<r\right>\right)^{-2}$ as $\left<r\right>$
approaches $L_{\rm C}$, just like the worm-like chain model.\cite{Marko95}

These results can be directly compared with the results derived by
MacKintosh and coworkers.\cite{MacKintoshPRL95} In their
entropic model, filaments continually undergo thermal bending
motions as they are subjected to an external tensile force. At each
instance and applied force the configurational state of the filament
changes, as a result of the interaction with the surrounding fluid. In
this {dynamic} model, each mode amplitude changes in
time and with force, {i.e.}, $a_n=a_n(t;\fc)$.  The free-energy functional
of the filament is written as\cite{MacKintoshPRL95}
\begin{equation}
\mathscr{H}=\frac{1}{2}\kappaf\int_{0}^{L_{\rm C}}\left(\frac{\d \theta}
{\d s}\right)^2\d s + \frac{1}{2}\fc\int_{0}^{L_{\rm C}}\theta^2(s)\d s,
\label{hamiltonian2}
\end{equation}
where the first term is the filament's bending energy and the second term is the 
work of contracting against the applied tension. Using this Hamiltonian, the force-dependence of 
the mean-squared value of $a_n$ can be found from equipartition at each force, and reads
\begin{equation}
\left<a_n^2\right>(\fc)=\frac{1}{L_{\rm P}}\frac{1}{q_n^2+\fc/\kappaf}.
\label{meansquaredan2}
\end{equation}
Then, by making use of Eq.~(\ref{fcdepeted}), the time- or ensemble-averaged end-to-end 
distance $\left<\tilde{r}\right>$ becomes\cite{rmrktilde}
\begin{equation}
\left<\tilde{r}\right>(\varphi)=L_{\rm C}-\frac{L_{\rm C}^2}{2\pi^2L_{\rm P}}
\sum_{n \ge 1}\frac{1}{n^2+\varphi},
\label{fcdepetedAVG5}
\end{equation}
and can be rewritten in the following form:
\begin{equation}
\left<\tilde{r}\right>(\varphi)=L_{\rm C}-\frac{L_{\rm C}^2}{4L_{\rm P}\pi^2\varphi}
\left(\pi \sqrt{\varphi}\coth\left[\pi \sqrt{\varphi}\right]-1\right).
\label{fcdepetedAVG6}
\end{equation}

\begin{figure}[t] % Fig. 6
\begin{center}
\includegraphics[width=0.45\textwidth]{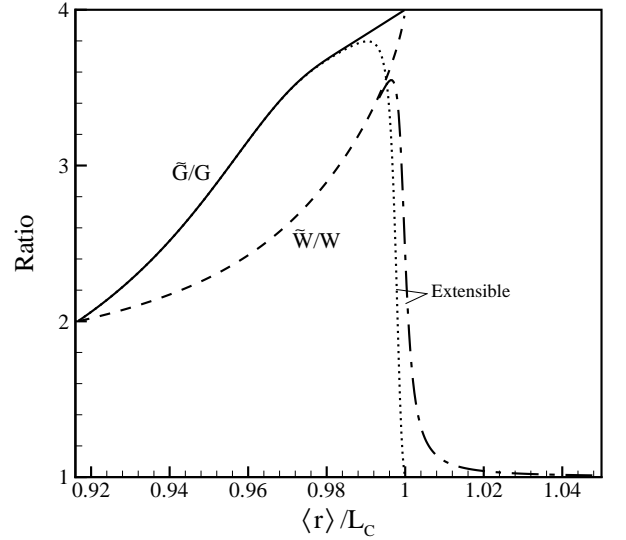}
\end{center}
\caption{The the evolution of the dynamic-to-static stiffness ratio ($\tilde{G}/G$) and external work ratio 
($\tilde{W}/W$) with$\left<r\right>/L_{\rm C}$. Results are plotted for inextensible filaments 
($\tilde{G}/G$ solid curve; $\tilde{W}/W$ dashed curve) and extensible filaments ($\tilde{G}/G$ dotted 
curve; $\tilde{W}/W$ dash-dotted curve). The parameters used for the calculation: $L_{\rm P}=L_{\rm C}=10~\mu$m, $\kappaf=8.53\times10^{-17}~\mbox{Nm}^2$ and $\mu=16~\mbox{N}$.}
\label{ratios}
\end{figure}

The dynamic behavior, shown in Fig.~\ref{feravg} by the dashed curve, is thus quite similar, 
but not equal to the behavior according to the static
description. The small-force response can be approximated by
\begin{equation}
\left<\tilde{r}\right>(\varphi) \approx \left(L_{\rm C}-\frac{L_{\rm C}^2}{12L_{\rm P}}\right) 
+ \frac{L_{\rm C}^2 \pi^2 \varphi}{180L_{\rm P}}.
\label{fcdepetedAVG7}
\end{equation}
so that the slack distance in the dynamic model is equal to that in
the static model, cf. Eq.~(\ref{fcdepetedAVG3}). However, 
the initial stiffness
\begin{equation}
\tilde{G}_{\rm b}^0=\frac{180L_{\rm P}\kappaf}{L_{\rm C}^{4}}
\label{bendingstiffness2}
\end{equation}
is a factor two larger than the static stiffness, cf. Eq.~(\ref{bendingstiffness1}). At large forces, 
the diverging response is also similar to what we found in the static model, but from 
Eq.~(\ref{fcdepetedAVG6}) the limiting behavior near full extension,
\begin{equation}
\left<\tilde{r}\right>(\varphi) \rightarrow L_{\rm C}-\frac{L_{\rm C}^2}{4L_{\rm P}\pi \sqrt{\varphi}},
\label{fcdepetedAVG8}
\end{equation}
has a stiffness that is a factor four higher than in (\ref{fcdepetedAVG4}). 
The dynamic-to-static stiffness ratio, $\tilde{G}/G$, is shown as a function of 
$\left<r\right>/L_{\rm C}$ in Fig.~\ref{ratios} (solid curve). As mentioned above, the ratio gradually 
increases from 2 at $\left<r\right>/L_{\rm C}=1-L_{\rm C}/(12L_{\rm P})$ to 4 in the limit 
$\left<r\right>/L_{\rm C}\rightarrow1$.

The equality of the slack distance in both models is simply due
to the fact that, in the absence of an external force, the Hamiltonian
used in the dynamic model, Eq.~(\ref{hamiltonian2}), reduces to the
bending energy, Eq.~(\ref{hamiltonian}), that is also used for
equipartition in the static model. However, as soon as the chain is
subjected to a tensile force (no matter how small in magnitude), the
static and dynamic mechanical responses deviate. This deviation is
caused by the difference in the chain's energy used for
equipartition. In the static model, only the initial undulations are
taken into account by equipartition at $\fc=0$. Next, each frozen-in
filament is subjected to a tensile force, thereby reducing the
internal bending energy of the system. In the dynamic model,
equipartition is imposed on the total free-energy functional, comprising 
both the bending energy and the chain's work of contracting against the 
applied force: on average each harmonic mode in the chain's total internal
energy remains $\kB T/2$ during pulling, even though its bending
component reduces. At each force, there must therefore be sufficient
time for a complete energy redistribution between different bending
modes and between the bending and tension energies, to ensure that the
chain's energy [Eq.~(\ref{hamiltonian2})] remains constant. Compared
to static filaments, a larger force is required to reach a certain
end-to-end distance in order to overcome the effect of dynamic
undulations, thus resulting in a higher stiffness.

Since energy is the origin of the difference in static and dynamic mechanical response, 
it is of interest to study the external work $W$ needed to reach an average end-to-end 
distance $\left<r\right>$, which is calculated as
\begin{equation}
W(\left<r\right>)=\frac{\pi^2 \kappaf}{L_{\rm C}^2}\int_{\left<r_0\right>}^{\left<r\right>}
\varphi(s)\d s
\label{work}
\end{equation}
with $\left<r_0\right>=L_{\rm C}-L_{\rm C}^2/(12L_{\rm P})$. By numerically inverting 
Eqs.~(\ref{fcdepetedAVG2}), (\ref{fcdepetedAVG6}) and by integration with Eq.~(\ref{work}), 
it can be shown that $W(\left<r\right>)$ diverges as $\left<r\right>$ approaches $L_{\rm C}$, 
both in the static and in the dynamic model. Therefore, the dynamic-to-static work ratio, $\tilde{W}/W$, 
approaches the same value as the stiffness ratio, {i.e.},
\begin{displaymath}
\mathop{\lim}\limits_{\left<r\right>\to L_{\rm C}}(\tilde{W}/W)=\mathop{\lim}
\limits_{\left<r\right>\to L_{\rm C}}(\tilde{G}/G)=4.
\end{displaymath}
The dependence of the work ratio on $\left<r\right>$ for inextensible filaments is shown in 
Fig.~\ref{ratios} (dashed curve) and indeed increases from 2 at $\left<r\right>=\left<r_0\right>$ 
to 4 near full extension, similar to the stiffness ratio. Compared to static chains, the energy 
needed to extend dynamically undulating chains close to their contour length is about four times 
larger. In both cases, however, the completely straight configuration will never be reached as a 
result of the energy divergence. 

The origin of this divergence is different in the two models and can
be explained by allowing thermal bending motions to excite only
the first order mode $a_1$ in the Fourier series (\ref{Fourier1}). The
average value of the end-to-end distance in the absence of force then is
$\left<r_0\right>=L_{\rm C}-L_{\rm C}^2/(2 \pi^2 L_{\rm P})$,
which follows directly from Eqs.~(\ref{fcdepetedAVG1}) and (\ref{fcdepetedAVG5}). 
The force-extension relations can be inverted analytically and the external work 
according to Eq.~(\ref{work}) can be readily calculated. For static chains, the 
energy needed for complete stretching is
\begin{displaymath}
\frac{\pi^2 \kappaf}{L_{\rm C}^2}\mathop{\lim}\limits_{\left<r\right>\to L_{\rm C}}
\int_{\left<r_0\right>}^{\left<r\right>}\left(-1+\sqrt{\frac{L_{\rm C}^2}{2\pi^2 
L_{\rm P}(L_{\rm C}-s)}}\right)\d s=\frac{\kB T}{2},
\end{displaymath}
and, thus, is equal to the average initial energy stored in bending. For a finite 
number of allowed modes, static chains can be stretched to their contour length by 
an energy input equal to the initially stored bending energy. The origin of the energy 
divergence for static chains is merely the result of allowing an infinitely large number 
of bending configurations in the Fourier series ($N \to \infty$). However, for dynamic 
chains the energy needed for full stretching is
 \begin{displaymath}
\frac{\pi^2 \kappaf}{L_{\rm C}^2}\mathop{\lim}\limits_{\left<r\right>\to L_{\rm C}}
\int_{\left<r_0\right>}^{\left<r\right>}\left(-1+\frac{L_{\rm C}^2}{2\pi^2 
L_{\rm P}(L_{\rm C}-s)}\right)\d s,
\end{displaymath}
and diverges. This is a result of the fact that equipartition is applied on the total 
free-energy functional [Eq.~(\ref{hamiltonian2})]. Hence, dynamic chains can never be pulled 
straight completely, even if only a single bending mode is allowed.

In case of extensible filaments we have to incorporate the axial stretching energy
\begin{equation}
\mathscr{H}_s=\frac{\mu}{2} \int_{0}^{L_{\rm C}+\Delta L_{\rm C}} \left(\frac{\d l}{\d s}\right) ^2\d s,
\label{hamiltonian3}
\end{equation}
where $\d l/\d s$ is the relative length change along the
filament and $\Delta L_{\rm C}$ the total increase in contour
length. Equation~(\ref{hamiltonian3}) determines the Hookean response
of the filament with a stretching stiffness $\mu$. The
ensemble-averaged mechanical behavior of such filaments can be found
by applying Eq.~(\ref{fcdepeted3}) to Eq.~(\ref{fcdepetedAVG1}) for
static chains and to Eq.~(\ref{fcdepetedAVG5}) for dynamic chains. The
results are shown in the inset of Fig.~\ref{feravg}, with
$\mu=16$~N. As $\left<r\right>$ approaches $L_{\rm C}$, the mechanical
response is clearly dominated by axial stretching of the
filaments. The energy stored in stretching dominates the total energy
and the deviation between the dynamic and static model vanishes around
$\left<r\right>=L_{\rm C}$. This can be shown even better by studying
the dynamic-to-static stiffness and work ratios for extensible filaments,
that are also shown in Fig.~\ref{ratios} (dotted and dash-dotted
curves, respectively). While following the curves for inextensible filaments 
at small end-to-end displacements (bending-dominated regime), both ratios decrease 
to one around $\left<r\right>=L_{\rm C}$, where the
stretching-dominated regime starts. 

The stiffening of individual filaments cannot be easily overestimated: 
the average axial stiffness increases from $G_{\rm b}=7.7~\mbox{N/m}$ for
static filaments, Eq.~(\ref{bendingstiffness1}), or $G_{\rm b}=15.4~\mbox{N/m}$ 
for dynamic filaments, Eq.~(\ref{bendingstiffness2}), in the bending-dominated 
regime, to a value of $G_{\rm s}=\mu/L_{\rm C}=1.6\times10^6~\mbox{N/m}$ in the 
stretching-dominated regime. Since $G_{\rm b}$ is six orders of magnitude smaller 
than $G_{\rm s}$, we will neglect from now on the entropic, axial stiffness 
due to internal bending. For an individual filament with a slack distance 
$\xi$, the dependence of the axial stiffness on the end-to-end displacement $u$ is 
thus approximated by
\begin{equation}
G_{1}(u,\xi;L_{\rm C})\approx \left\{\begin{array}{ll}
0 & \mbox{for } u<\xi\\
{\mu}/{L_{\rm C}} & \mbox{for } u\ge\xi\\
\end{array}.\right.
\label{g1}
\end{equation}
Since the probability of a filament having a slack distance between $\xi$ and 
$\xi+\d \xi$ is equal to $\mathscr{G}(L_{\rm C}-\xi;L_{\rm C};L_{\rm P})\d \xi$ 
[cf. Eq.~(\ref{rdf1})], the average stiffness of an ensemble of chains
with different slacks can be expressed by
\begin{eqnarray}
\left<G_{1}\right>(u;L_{\rm C};L_{\rm P})=\int_{0}^{L_{\rm C}}\mathscr{G}(L_{\rm C}-\xi;
L_{\rm C};L_{\rm P})G_{1}(u,\xi;L_{\rm C})\d \xi \nonumber\\
\approx \frac{\mu}{L_{\rm C}}\int_{0}^{u}\mathscr{G}(L_{\rm C}-\xi;L_{\rm C};L_{\rm P})\d \xi.~~
\label{g1avgmod}
\end{eqnarray}
This result can be integrated to yield the average force-extension relation:
\begin{equation}
\left<f_{c}\right>(u;L_{\rm C};L_{\rm P})=\int_{u'=0}^{u}\left<G_{1}\right>(u';L_{\rm C};L_{\rm P})\d u'
\label{fcavgmod}
\end{equation}
It is important to note the difference in averaging procedures used
above: in Eqs.~(\ref{fcdepetedAVG1}) and (\ref{fcdepetedAVG5}) the
average is taken over all end-to-end lengths at constant force,
whereas in Eqs.~(\ref{g1avgmod}), (\ref{fcavgmod}) the average is
taken over all forces at constant displacement. 

\begin{figure}[t] %Fig. 7
\begin{center}
\includegraphics[width=0.47\textwidth]{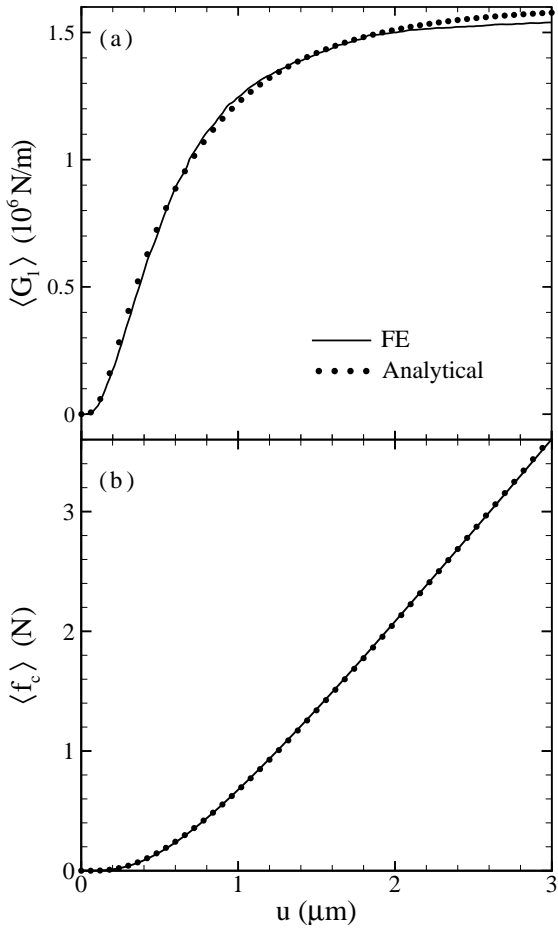}
\end{center}
\caption{{(a)} The ensemble-averaged stiffness
$\left<G_1\right>(u;L_{\rm C};L_{\rm P})$ of individual semiflexible filaments with
$L_{\rm P}=L_{\rm C}=10~\mu$m, and {(b)}, the corresponding
average force-extension relation $\left<f_{c}\right>(u;L_{\rm C};L_{\rm P})$. Results are
shown of the analytical calculations using Eqs.~(\ref{g1avgmod}),
(\ref{fcavgmod}) (closed circles), and from the finite-element (FE)
calculations (drawn curve). The mechanical parameters used are:
$\mu=16~\mbox{N}$ and $\kappaf=8.53\times 10^{-17}~\mbox{Nm}^2$ 
(the latter value is needed in the FE calculation only).} 
\label{onefilavg}
\end{figure} 

The average stiffness, $\left<G_1\right>(u;10~\mu\mbox{m};10~\mu\mbox{m})$ according to
Eq.~(\ref{g1avgmod})  is plotted in Fig.~\ref{onefilavg}(a) [solid
circles]. As can be seen, the average stiffness gradually increases
from 0 for small displacements to a saturation value of $\mu/L_{\rm C}$
for large displacements. This enthalpic stiffening thus results from
an increase in the number of filaments that are stretched beyond their slack
distance. The corresponding average force-extension relation is shown in 
Fig.~\ref{onefilavg}(b) [solid circles].

To investigate and check the mechanical response found in the
analytical model, we have performed 2D finite-element (FE) calculations on
individual filaments ($\mu=16~\mbox{N}$, $\kappaf=8.53\times10^{-17}~\mbox{Nm}^2$). 
Filaments were generated as described in Section II using 
Eqs.~(\ref{Fourier1}) and (\ref{standarddeviation}) with $L_{\rm
P}=L_{\rm C}=10~\mu$m, fixed at one end and pulled by prescribing the
displacement of the other end along the direction of the initial
end-to-end vector.  The force-extension relation was calculated for
2000 randomly generated filaments, from which the ensemble-averaged
response was determined by averaging the force at constant
displacement; the stiffness was computed from
$\d \left<f_{c}\right>/\d u$. The results are shown by the solid
curves in Fig.~\ref{onefilavg}. As can be seen, the FE-predictions
correspond very well with the analytical model. What cannot be seen,
because of the plotting scale, is that
the initial average stiffness is $10^2~\mbox{N/m}$ instead of the value
$7.7~\mbox{N/m}$ calculated from Eq.~(\ref{bendingstiffness1}). This is due to
the difference in averaging procedure as explained above; if we
average over displacement at constant force, the result for small
displacements indeed corresponds to Eq.~(\ref{bendingstiffness1}).

\section{Affine model for elasticity in semiflexible biopolymer networks}

The ensemble-averaged force-extension relation of individual filaments
derived in Sect. III, enables us to  calculate the
mechanical response of cross-linked networks of such
filaments under simple shear. To do so, we follow the procedure used
by Wu and Van der Giessen for the elasticity of rubber networks.\cite{Wu93,Wu95} 

\begin{figure}[t] %Fig. 8
\begin{center}
\includegraphics[width=0.37\textwidth]{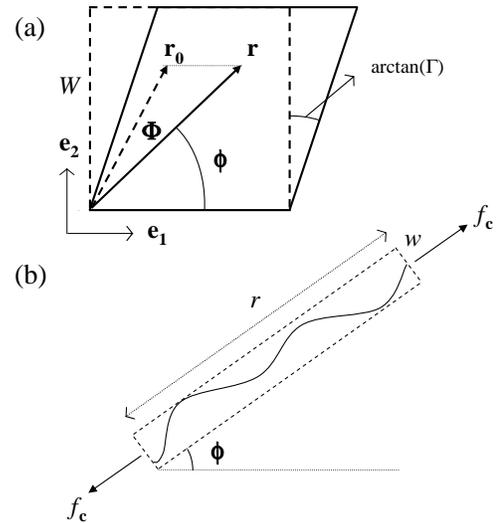}
\end{center}
\caption{{\bf (a)} Schematic representation of affine deformation of a two-dimensional 
unit cell under simple shear at a  strain $\Gamma$. The initial end-to-end vector ${\bf r_0}$ 
at an angle $\Phi$ with ${\bf e}_1$ has transformed to a vector ${\bf r}={\bf F}{\bf r_0}$ 
at an angle $\phi$. {\bf (b)} An (undulated) filament with end-to-end length $r$ at an 
angle $\phi$ under a tensile force $\fc$ in the deformed state. The area $rw$ covered 
by the filament is, on average, constant under the assumption that the network is incompressible.}
\label{affine}
\end{figure}

We consider a square unit cell of dimension $W$ spanned by base vectors ${\bf e}_1$ 
and ${\bf e}_2$ as indicated in Fig.~\ref{affine}(a). In the initial state, the unit 
cell contains $\Nf$ filaments of length $L_{\rm C}=L_{\rm P}=10~\mu$m, at random 
orientations. The areal and line density of filaments in the network are 
defined as $n=\Nf/W^2$ and $\rho=nL_{\rm C}$, respectively.  The angle between the initial filament's 
end-to end-vector ${\bf r_0}=r_0{\bf m_0}$ and ${\bf e}_1$ is denoted by $\Phi$. 

\begin{figure}[t] % Fig. 9
\begin{center}
\includegraphics[width=0.47\textwidth]{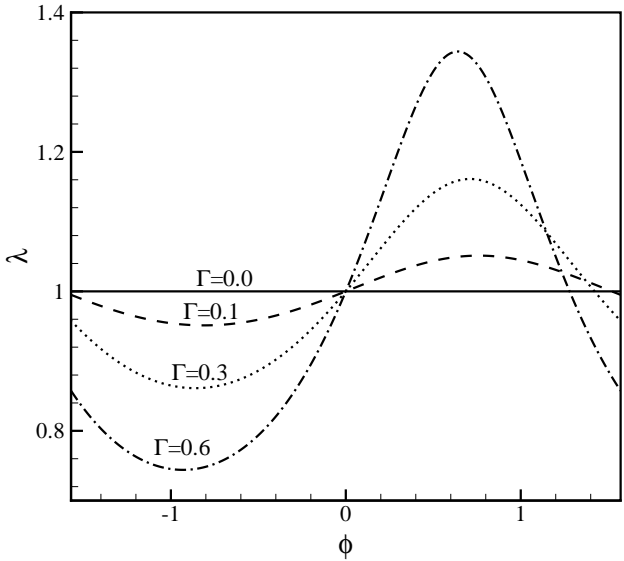}
\end{center}
\caption{The stretch $\lambda=r/r_0$ of a filament as a function of its orientation 
$\phi \in [-\pi/2,\pi/2]$ in the deformed state at various levels of
the shear $\Gamma$
according to Eq.~(\ref{stretch2}) for affine deformation.} 
\label{affinestretch}
\end{figure}

Next, the cell is subjected to a simple shear of strain  $\Gamma$, by
fixing the lower boundary and displacing the top boundary over a
distance $W\Gamma$, see Fig.~\ref{affine}(a). This deformation process is 
represented by the deformation gradient tensor ${\bf F}$ with components
\begin{equation}
\left[F_{ij}\right] = 
\left[ \begin{array}{cc}
1 & \Gamma \\
0 & 1 \\
\end{array} \right].
\label{Ftensor}
\end{equation}
Note that the shear deformation process satisfies $\det{\bf F}=1$,
expressing incompressibility of the filamental network.\cite{rmrkvolconservation} When it is assumed that the filaments deform in an 
{affine} manner, each filament's end-to-end vector ${\bf r_0}$ transforms to 
the vector ${\bf r}=r{\bf m}$ according to ${\bf r}={\bf F}{\bf r_0}$, as sketched in 
Fig.~\ref{affine}(a). Filaments thus rotate from $\Phi$ to $\phi$ and undergo a stretch 
of $\lambda=r/r_0$ that can be obtained from ${\bf F}$ through
\begin{equation}
\lambda^{-2}={\bf m}\left({\bf FF}^{\rm T}\right)^{-1}{\bf m},
\label{stretch}
\end{equation}
where ${\bf m}={\bf r}/r=\cos\phi{\bf e}_1 + \sin\phi{\bf e}_2$ is the
unit vector along the filament's end-to-end direction in the current,
deformed state. From Eqs.~(\ref{Ftensor}) and (\ref{stretch}) the stretch $\lambda$ can be 
expresssed in terms of $\phi$ and $\Gamma$ as
\begin{equation}
\lambda(\phi;\Gamma)=\sqrt{\frac{1+\tan^2\phi}{1-2\Gamma\tan\phi+(1+\Gamma^2)\tan^2\phi}}.
\label{stretch2}
\end{equation}
A filament that is at an angle $\phi$ in the current
state of shear strain $\Gamma$ is thus subjected to a
stretch $\lambda$ that is given by
Eq.~(\ref{stretch2}). Figure~\ref{affinestretch} shows the dependence
of $\lambda$ on $\phi$ for various strains $\Gamma$. 

\begin{figure}[t] % Fig. 10
\begin{center}
\includegraphics[width=0.45\textwidth]{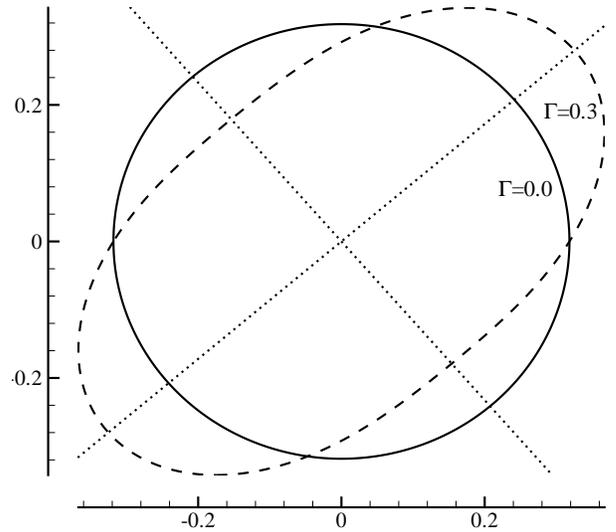}
\end{center}
\caption{Polar plot of the chain orientation distribution function $C(\phi;\Gamma)$ [CODF], at $\Gamma=0.0$ 
(solid curve) and $\Gamma=0.3$ (dashed curve). The dotted lines indicate the principal directions at 
$\Gamma=0.3$.} 
\label{codfplot}
\end{figure}

During shear, filaments also rotate and thus change their orientation. If $C(\phi;\Gamma)\d \phi$ 
is the probability of a filament with an orientation between $\phi$ and $\phi+\d \phi$ in the 
deformed state at strain $\Gamma$, it can be shown that $C(\phi;\Gamma)$ follows directly from 
the stretch $\lambda$ by~\cite{rmrkcodf1}
\begin{equation}
C(\phi;\Gamma)=C_0\lambda^2(\phi;\Gamma), 
\label{codf}
\end{equation}
where the normalization value $C_0=1/\pi$ is the initial, uniform probability distribution. This 
probability function $C(\phi;\Gamma)$ is also known as the ``chain orientation distribution function'' 
(CODF).\cite{Wu93,Wu95} Figure \ref{codfplot} shows a polar plot of the CODF for $\Gamma=0$ and 0.30 
in the angular range $[0,2\pi]$.\cite{rmrkcodf2} For $\Gamma=0$ 
there is no deformation and filaments are randomly distributed in the network, indicated by the spherical 
distribution. As the network deforms, filaments rotate in the straining direction, resulting in a 
nonspherical distribution, its anisotropy growing with increasing strain. For $\Gamma=0.3$ the principal 
directions for which a minumum and maximum stretch is obtained, are indicated by the mutually orthogonal, 
straight dashed lines. For this strain, filaments at an angle of $\phi=40.7$\deg undergo a maximum 
stretch of $\lambda=1.16$ and filaments at an angle of $\phi=-49.3$\deg~ a maximum compression 
($\lambda=0.86$).

Under the assumption of affine deformation, the 
Eqs.
(\ref{g1},\ref{g1avgmod},\ref{fcavgmod},\ref{stretch2},\ref{codf}) 
can be used to calculate the network response to simple shear. We start 
by considering an individual filament at an angle $\phi$ subjected to a stretch $\lambda(\phi,\Gamma)$ 
in the deformed unit cell of strain $\Gamma$, as sketched in Fig.~\ref{affine}(b). Assuming the network 
is incompressible, the area per chain, $rw$, remains constant, and is thus equal to $1/n$. The 
force acting on the filament, $f_{\rm c}$, can be tranformed to the
Cauchy traction $\sigma_{\rm c}$ acting on the continuum 
in which the filament is embedded, $\sigma_{\rm c}=f_{\rm c}/(bw)$, where $w$ is the 
cross-sectional width and $b$ is the unit out-of-plane length the force acts on. Similar to the approach 
by Wu and Van der Giessen\cite{Wu95} we define the {micro-stress} tensor  
$\bm{\sigma}_{\rm c}$ by
\begin{equation}
\bm{\sigma}_{\rm c}=\sigma_{\rm c}\left({\bf m}\otimes{\bf m}\right)-p{\bf I},
\label{microstress}
\end{equation}
with $\sigma_{\rm c}=n\lambda(\phi;\Gamma)f_{\rm c}r_0b^{-1}$ since $r=\lambda r_0$. In Eq.~(\ref{microstress}) 
the hydrostatic stress $p$ is included to account for the incompressibility of the network. The micro-stress tensor 
is the contribution of a single filament 
oriented along $\textbf{m}(\phi)$ to the stress of the network. Finally, with 
the areal density of chains $\d n$ having an orientation between $\phi$ and $\phi+\d \phi$ given by
\begin{equation}
\d n=nC(\phi;\Gamma)\d \phi, 
\end{equation}
the overall or {macro-stress} tensor $\bm{\sigma}$ of the network can be evaluated by the average of the 
micro-stress tensor $\bm{\sigma}_{\rm c}$ over the individual chains in the network, {i.e.},
\begin{equation}
\bm{\sigma}=\frac{1}{n}\int\bm{\sigma}_{\rm c}\d n.
\label{macrostress}
\end{equation}
\begin{figure}[t] % Fig. 11
\begin{center}
\includegraphics[width=0.47\textwidth]{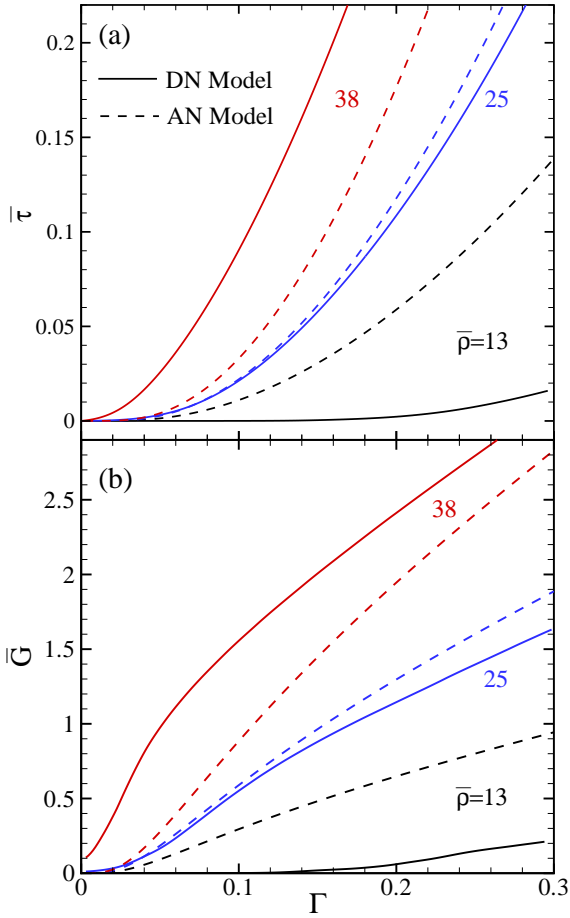}
\end{center}
\caption{{(a)} Shear stress $\overline{\tau}$ versus shear strain $\Gamma$ response for biopolymer 
networks comprising filaments with $L_{\rm P}=L_{\rm C}=10~\mu$m, $\kappaf=8.53\times 10^{-17}$~Nm$^2$ 
and $\mu=16$~N, and {(b)}, the corresponding shear stiffness $\overline{G}=\textrm{d}\overline{\tau}/
\textrm{d}\Gamma$. Results are plotted for three filament densities: $\overline{\rho}$=13 (black curves), 
25 (blue) and 38 (red). The solid curves display the results from the discrete-network (DN) calculations, 
the dashed curves are the results from the affine-network (AN) model, Eq.~(\ref{macroshearstressnorm}).} 
\label{nwresponse1}
\end{figure}
\begin{figure*}[t!] % Fig. 12
\begin{center}
\includegraphics[width=\linewidth]{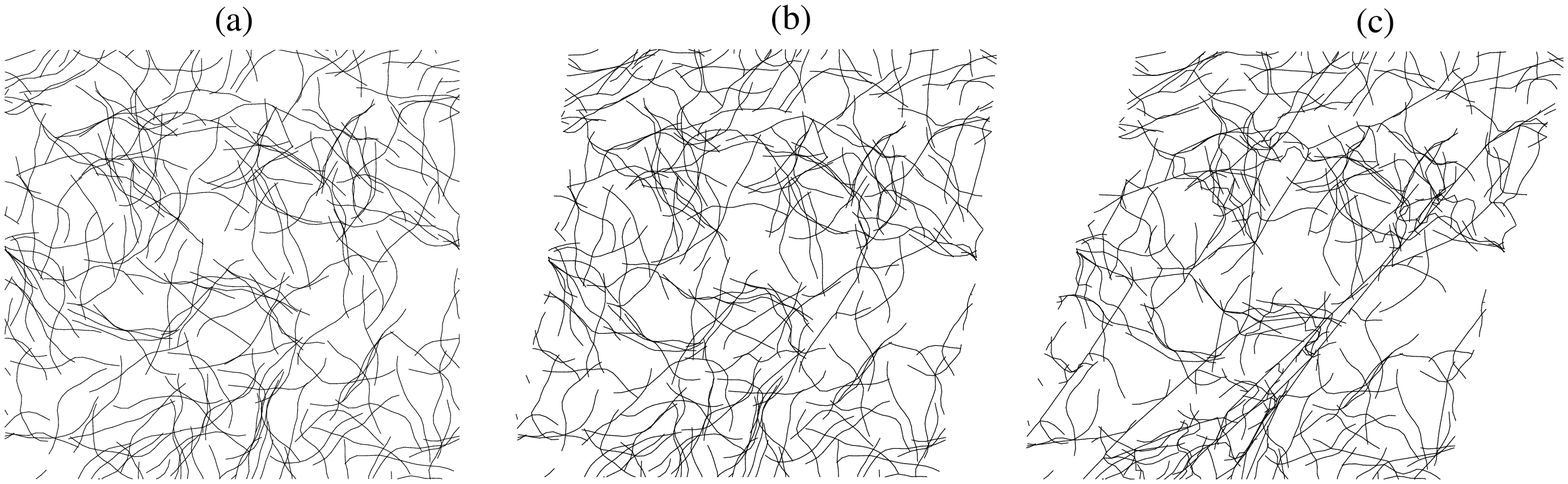}
\end{center}
\caption{Shear deformation of the discrete network (DN). (a) Initial, $\Gamma=0$; (b) intermediate, $\Gamma=0.125$
and (c) large strain, $\Gamma=0.285$, network configurations for a randomly generated network of density 
$\overline{\rho}=13$, with $W=4L_{\rm C}=40~\mu\mbox{m}$.} 
\label{networks}
\end{figure*}
It is important to note here that since the force acting on an individual filament depends on its end-to-end 
displacement, it therefore depends on its current orientation $\phi$ and applied shear strain $\Gamma$ through 
$f_{\rm c}(\phi;\Gamma)=f_{\rm c}(u(\phi;\Gamma))=f_{\rm c}(r_0(\lambda(\phi;\Gamma)-1))$. 

We assume that the filaments in the network only stretch and compress upon mechanical loading, 
but do not bend/buckle. In Sect. III we showed that stretching by internal bending can also be neglected, 
and that axial stretching with stiffness $\mu/L_{\rm C}$ dominates the ensemble-averaged single-filament force-extension 
relation.  The angular region in which stretching occurs ($\lambda \ge 1$) follows directly from 
Eq.~(\ref{stretch2}) and is given by $\phi \in [0,\arctan(2\Gamma^{-1})]$. In this region filaments experience 
an elongation of their end-to-end distance; filaments
outside this region are under compression ($\lambda \le 1$) and deform by soft (internal) bending and contribute 
negligibly to the overall network stiffness. Hence, we adopt the force-extension relation based on 
Eqs.~(\ref{g1})--(\ref{fcavgmod}) in the angular region $\phi \in [0,\arctan(2\Gamma^{-1})]$. The dimensionless shear stress, $\overline{\tau}=\sigma_{12} b L_{\rm C}/\mu$, from 
Eqs.~(\ref{codf})--(\ref{macrostress}) can then be written as
\begin{equation}
\overline{\tau}=\frac{\overline{\rho}}{\pi \mu}\frac{\left<r_0\right>}{L_{\rm C}}\int\limits_{0}^{\arctan(2\Gamma^{-1})}\lambda^3(\phi;\Gamma)\left<f_c\right>(\phi;\Gamma)\sin\phi\cos\phi \d \phi,
\label{macroshearstressnorm}
\end{equation}
where $\overline{\rho}=\rho L_{\rm C}=\Nf\left(L_{\rm C}/W\right)^2$ is a dimensionless filament density. 
Figure~\ref{nwresponse1}(a) presents the calculated $\overline{\tau}$
versus $\Gamma$-curves for
three densities $\overline{\rho}=$13, 25 and 38 (dashed curves). The corresponding shear stiffness, 
$\overline{G}=\textrm{d}\overline{\tau}/\textrm{d}\Gamma$, is shown in Fig.~\ref{nwresponse1}(b). As can be seen, 
stiffening of the network is indeed observed, resulting from the average stiffening of individual filaments 
in the network (Fig.~\ref{onefilavg}), or said differently, stiffening
results from an increase in the fraction of filaments 
that are stretched beyond their slack distance.

These results can be directly compared to 2D numerical discrete-network (DN) calculations using the finite-element 
method.\cite{OnckPRL05} Filaments of contour length $L_{\rm C}$ and persistence length $L_{\rm P}$ are 
randomly placed into a square cell of dimension $W$ at random orientations, with proper account of periodicity.  
Each filament is discretized into 15 equal-sized, Euler-Bernoulli beam
elements that account for  stretching 
(with axial stiffness $\mu$) and bending (with stiffness $\kappaf$). In accordance with Sect. II, 10 Fourier modes are used to 
describe the initial shape of the undulated filaments. The points were filaments overlap are considered to be stiff 
cross-links, where the rotation and the displacement of the two filaments at the cross-link is equal. 
Figure~\ref{networks}(a) shows an example of a randomly generated network of density $\overline{\rho}=13$ 
that is in its initial, stress-free configuration. Filaments crossing either the upper or lower boundary of the cell 
are perfectly bonded to rigid top and bottom plates, respectively. Next, the top plate is displaced horizontally relative 
to the bottom plate over a distance $\Gamma W$, corresponding to an applied shear strain of $\Gamma$. Geometry changes 
are accounted for by an updated Lagrangian finite-strain
formulation. Figures~\ref{networks}(b) and (c) show the 
network of Fig.~\ref{networks}(a) in the deformed state at shears of $\Gamma$=0.125 and 0.285, respectively. 
The macroscopic shear stress $\tau$ is calculated from the cumulative, horizontal reaction force of nodes at the 
top plate, divided by the cell width $W$. Convergence studies were performed to ensure that the results are not 
affected by the number of elements per filament and the cell size $W$.

\begin{figure}[b] % Fig. 13
\begin{center}
\includegraphics[width=0.45\textwidth]{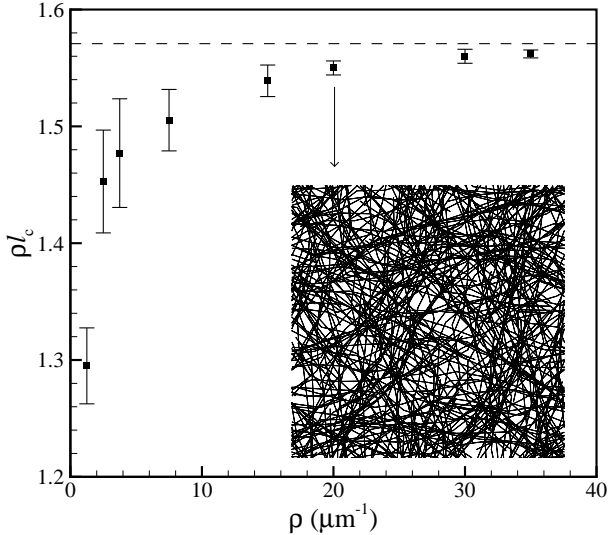}
\end{center}
\caption{The product $\rho \lc$ versus the filament network density $\rho$ (in $\mu\mbox{m}^{-1}$), determined from 
and averaged over randomly generated networks (solid squares). The dashed line indicates the limiting, high-density 
value of $\pi/2$. The inset shows a random network at a density of $\rho=20~\mu\mbox{m}^{-1}$($\overline{\rho}=200$), 
with $W=L_{\rm C}=10~\mu\mbox{m}$.}
\label{rhodeplc}
\end{figure}

The stress-strain response (averaged over 10 different, random realizations) at three network filament 
densities of $\overline{\rho}=$13, 25 and 38 is included in Figure~\ref{nwresponse1} (solid curves). It can 
be seen that, at a relatively small density ($\overline{\rho}=$13), the discrete-network (DN) calculations 
result in a much softer response compared to the affine-network (AN)
calculations. Closer examination of the DN results reveals, first, that
the small-strain mechanical response is dominated by bending and buckling of 
filaments,\cite{OnckPRL05} thus yielding a soft response.  Second, at somewhat larger strains, filaments perform 
additional, nonaffine motions during shearing: filaments reorient themselves in the direction of 
straining by rotations and translations. These nonaffine, geometrical network reorientations can 
be clearly seen by comparison of Fig.~\ref{networks}(b) with Fig.~\ref{networks}(a). Both effects, bending/buckling of 
filaments and local network reorientations, are not taken into account in the AN model, where instead, 
at small strains, all filaments with an orientation between 0 and $\pi/2$ are subjected to stretch. The 
response in the AN model is therefore dominated by stretching of filaments, contrary to what is observed in 
the DN computations. However, as the strain increases, more filaments become oriented in the direction of 
straining, as seen in Fig.~\ref{networks}(c): strings of stretched filaments connect the top and 
bottom plate of the cell. Thus, at larger strains, the response
according to the DN calculations is 
also dominated by stretching of filaments, which explains the stiffening in the DN curves in Fig.~\ref{nwresponse1} 
at low densities. Since the relative amount of filaments in stretching remains small compared to that of the 
AN model, the stiffness determined from the DN calculations remains small compared to the stiffness from the 
AN model (black curves in Fig.~\ref{nwresponse1}). 

In contrast to small densities, at larger densities 
(e.g., $\overline{\rho}=38$) the DN calculations exhibit a higher stiffness compared to the AN calculations 
(red curves in Fig.~\ref{nwresponse1}). The reason for this is that the cross-link density increases with
filament density, with two effects. First, the higher cross-link
density results in a more rigid network structure in which each filament is cross-linked 
to more filaments, thereby decreasing the freedom of filaments to undergo nonaffine reorientations. The 
transition from a bending-dominated regime to a stretching-dominated regime therefore shifts to 
smaller strains as can be seen by comparing the solid DN curves in Fig.~\ref{nwresponse1}. Second, 
chain segments between cross-links contain less slack than the original filaments: segments are 
thus stretched beyond their slack distance at smaller strains leading to a higher stiffness. 
In the AN model, this reduction in slack is not taken into account, but cross-links are 
assumed to be at the ends of the filament, independent of the density of filaments in the network.
Density enters the AN model only through the prefactor in Eq.~(\ref{macroshearstressnorm}), which
explains the (self-) similar shape of the AN curves in Fig.~\ref{nwresponse1}. 

\begin{figure}[t] % Fig. 14
\begin{center}
\includegraphics[width=0.47\textwidth]{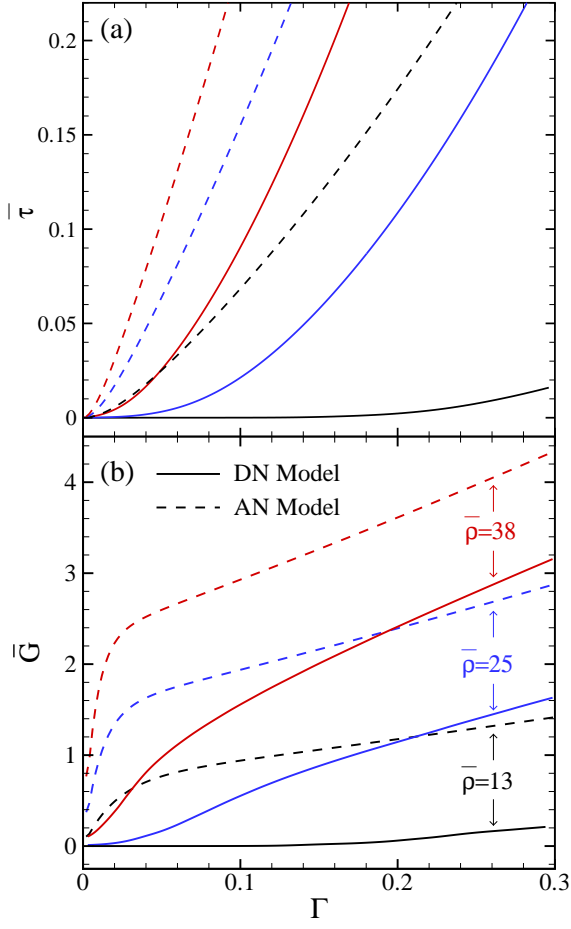}
\end{center}
\caption{{(a)} Shear stress $\overline{\tau}$ versus shear strain $\Gamma$ response for biopolymer 
networks comprising filaments with $L_{\rm P}=L_{\rm C}=10~\mu$m, $\kappaf=8.53\times 10^{-17}$~Nm$^2$ 
and $\mu=16$~N, and {(b)}, the corresponding shear stiffness $\overline{G}=\textrm{d}\overline{\tau}/
\textrm{d}\Gamma$. Results are plotted for three filament densities: $\overline{\rho}$=12.5 (black curves), 
25 (blue) and 37.5 (red). The solid curves display the results from the DN calculations (same as in Fig.~\ref{nwresponse1}), the dashed curves are the results from the density-corrected AN model.} 
\label{nwresponse2}
\end{figure}

To correct for the decrease in slack in the AN calculations, we 
first have to determine the {average} cross-link distance, $\lc$, in the network as a 
function of the density $\rho$. The values obtained from a series of
generated networks are shown in Fig.~\ref{rhodeplc} as the 
product $\rho\lc$ versus $\rho$. The product gradually increases from $1.2950\pm0.0325$ at 
$\rho=1.25~\mu\mbox{m}^{-1}$ ($\overline{\rho}=12.5$) to $1.5618\pm0.0034$ at $\rho=35.0~\mu\mbox{m}^{-1}$ 
($\overline{\rho}=350$) and thus approaches the limiting, 2D
theoretical value of 
$\pi/2$.\cite{rmrkproduct,HeadPRE03rc} The inset shows an undeformed network with $W=L_{\rm C}=10~\mu\mbox{m}$ 
at a high density of $\rho=20.0~\mu\mbox{m}^{-1}$, exemplifying the short average distance between the cross-links 
($\lc=0.0775\pm0.0003~\mu\mbox{m}\ll L_{\rm C}$). Next, the distribution in end-to-end lenghts $r_0$ of segments 
at a density $\rho$ is given by $\mathscr{G}(r_0;\lc(\rho);L_{\rm P})$ from Eq.~(\ref{rdf1}), with $L_{\rm C}$ 
replaced by $\lc(\rho)$. From this, the ensemble-averaged mechanical behavior of a segment of mean length $\lc(\rho)$,
{i.e.}, $\left<G_{1}\right>(u;\lc(\rho);L_{\rm P})$ and $\left<f_{c}\right>(u;\lc(\rho);L_{\rm P})$, 
follows directly from Eqs.~(\ref{g1avgmod})--(\ref{fcavgmod}). This then serves as input for the density-corrected 
AN response $\overline{\tau}$ given by Eq.~(\ref{macroshearstressnorm}).\cite{rmrkprefactor}
The corrected AN response is plotted in Fig.~\ref{nwresponse2} for three densities $\overline{\rho}=12.5,~25 
\mbox{ and } 37.5$ (dashed curves). For comparison, the figure also displays the DN calculations from Fig.~\ref{nwresponse1} 
(solid curves). As can be directly seen, all AN curves have gone up due the reduction in slack: already at small strains
filamental segments are stretched beyond their slack distance resulting in a higher stress and stiffness. After the density 
correction, all AN calculations exhibit a higher stiffness than the DN calculations, irrespective of the density: as said before,
this can be attributed to nonaffine network deformations (bending/buckling, local network reorientations) in the DN calculations 
that are not taken into account in the AN model.

\begin{figure}[t] % Fig. 15
\begin{center}
\includegraphics[width=0.47\textwidth]{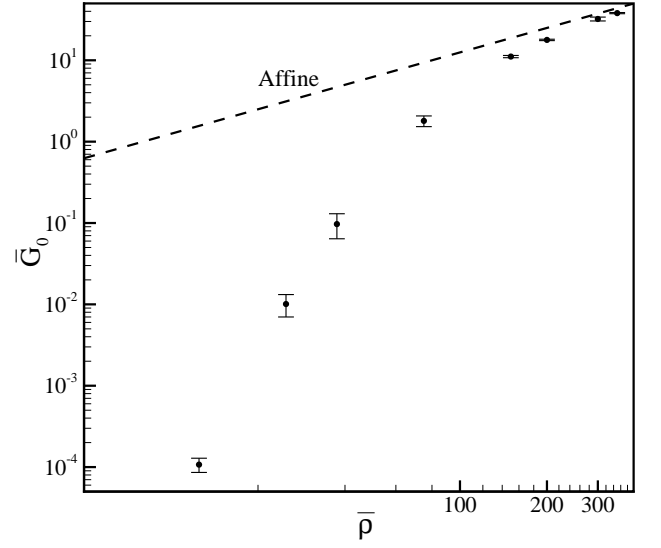}
\end{center}
\caption{Initial, small-strain stiffness $\overline{G}_0$ versus the density $\overline{\rho}$, calculated using 
the DN model (solid squares). The values of $\overline{G}_0$ result from an average over 10 random networks in 
case $\overline{\rho}<100$ and 5 networks for higher densities. The dashed line is the affine upper limit according 
to Eq.~(\ref{affinitstiff})} 
\label{upperbound}
\end{figure}

In calculating the density-corrected response, the angular integration was performed in the interval 
$[0,\arctan(2\Gamma^{-1})]$, see Eq.~(\ref{macroshearstressnorm}), assuming that segments still contain a considerable amount of slack, 
resulting in a very compliant response in case such a segment is under axial compression. However, 
as discussed above, when the filament density increases, segments
between cross-links become shorter and virtually straight. For a
network consisting of only straight filaments, segments under
compression exhibit the same axial stiffness as in tension, i.e.,
$\mu/\lc$, corresponding to an axial force of $(\mu/\lc)u$. Hence, the integration 
should be performed over the entire angular range, $[0,\pi]$, and with the segmental force equal to 
$\mu[\lambda(\phi;\Gamma)-1]$, the network response is evaluated as
\begin{equation}
\overline{\tau}=\frac{\overline{\rho}}{\pi}\int\limits_{0}^{\pi}\lambda^3(\phi;\Gamma)(\lambda(\phi;\Gamma)-1)\sin\phi\cos\phi \d \phi.
\label{macroshearstressnorm2}
\end{equation}
Eq.~(\ref{macroshearstressnorm2}) thus gives the limiting, affine mechanical behavior of biopolymer networks of high density. 
Whether or not filaments are undulated and contain slack is not relevant at this point, since the segments of 
average length $\lc \ll L_{\rm C}$ are straight at high densities [see inset Fig.~(\ref{rhodeplc})]. As we showed before, the DN
calculations exhibited nonaffine deformation characteristics at low densities. However, at higher densities the DN calculations are
also expected to show an affine behavior. To check this, we have investigated the initial, small-strain network stiffness, 
$\overline{G}_0$, as a function the density $\overline{\rho}$, see Fig.~\ref{upperbound} (solid squares). 
The affine value follows directly from the derivative of Eq.~(\ref{macroshearstressnorm2}) by a first-order Taylor expansion 
of the integrand with respect to $\Gamma$, and equals\cite{expansion}
\begin{equation}
\overline{G}_{0,\rm aff}=\frac{\overline{\rho}}{8}.
\label{affinitstiff}
\end{equation}
Note that, using the high-density limit $\rho\lc=\pi/2$,\cite{rmrkproduct} the stiffness can be rewritten as 
$G_{0,\rm aff}=\pi \mu/(16\lc)$, corresponding to the result found by
Head \textit{et al.}.\cite{HeadPRL03} The normalized 
affine limit, $\overline{G}_{0,\rm aff}$, is included in
Fig.~\ref{upperbound} as the dashed line. It can be 
seen that, as the density increases, the DN calculations indeed approach the AN limit. At a density of $\overline{\rho}=350$ 
the DN calculations result in a stiffness of $\overline{G}_0=38.0\pm0.6$, close to the affine value of 43.75.

Finally, it is interesting to see that in the networks' stretching regime, {i.e.}, at large strains, the network 
stiffness (both AN and DN) continues to increase despite the fact that the individual segments have a constant axial stiffness 
$\mu/\lc$ in stretching [see Fig.~\ref{nwresponse2}(b)]. From this one would expect the stiffness to converge to a fixed 'steady-state' 
value. However, due to nonlinear geometrical effects at large strains, the stiffness increases with straining according to
\begin{equation}
\overline{G} \propto \overline{\rho}~\Gamma^2(1+\Gamma^2)^{-3/2},
\label{Gscaling}
\end{equation}
as derived in Appendix D. From further investigation of Eq.~(\ref{macroshearstressnorm2}) it follows that the stiffness starts 
to level off at very large, experimentally irrelevant strains of $\Gamma \approx 1.5$.

\section{Concluding remarks}

\noindent
We have performed a detailed investigation of the mechanical stiffening behavior of 2D semiflexible biopolymer filaments and networks of such filaments under simple shear. Such polymers undergo thermally excited bending motions, brought about by collisions with molecules of the surrounding fluid in which the filament is immersed. For inextensible filaments subjected to a tensile force, these undulations can be pulled out at the cost of an external energy input, resulting in an axial stiffness that can be calculated for two specific cases. 
First, we can assume no undulation dynamics to take place during tension. In this case, the filament is isolated from its surrounding fluid and the random, undulated state present in the filament is subsequently pulled out when subjected to a tensile force. In this {static} case, the axial stiffness results from internal bending with bending stiffness $\kappaf$; using equilibrium mechanics, the force-extension relation can be calculated and averaged over an ensemble of filaments. Second, undulation dynamics can be taken into account during pulling. By applying equipartition on the filament's total energy functional at each force, the ensemble-averaged, {dynamic} force-extension relation can be derived. This so-called entropic stiffening was previously derived by MacKintosh and coworkers.\cite{MacKintoshPRL95}

The resulting average mechanical response of single filaments in these
two cases turns out to be very similar. At small forces and end-to-end
displacements, both responses scale with $\kappaf^2$, with the initial
dynamic stiffness being a factor 2 larger than the static one. Close
to full stretching, as the average end-to-end length $\left<r\right>$
approaches the contour length $L_{\rm C}$, the force diverges as $\fc
\propto \left(L_{\rm C}-\left<r\right>\right)^{-2}$ in both models,
characteristic for the worm-like chain model, with the dynamic/static
stiffness ratio approaching 4. This indicates that the static and
dynamic approaches  are qualitatively very similar, capturing the same
physical dependence on the mechanical and geometrical parameters
$\kappaf,~L_{\rm P} \mbox{ and } L_{\rm C}$, but differ quantitatively
by a factor 2 to 4. 

In case of {extensible} filaments of stretching stiffness $\mu$,  enthalpic axial elongation dominates the filaments' mechanical behavior close to full stretching and both models thus  exhibit a quantitatively identical mechanical response. In addition, we have shown that for 2D filaments, one can very well describe the ensemble-averaged force-extension relation by completely neglecting contributions from internal bending and by merely taking into account the axial stretching of filaments and the distribution in slack.

Two-dimensional, cross-linked networks that have these semiflexible
filaments as building blocks stiffen under an applied shear
strain. Often, stiffening is attributed to the entropic stiffening of
individual filaments undergoing affine deformations. For the 2D
networks under investigation, we conclude by comparing analytical,
affine network (AN) calculations with discrete (finite-element)
network (DN) calculations that this cannot be the case. For networks
of low filament density, the mean cross-link distance $\lc$ is
relatively large and the resulting segments between cross-links contain
a relatively large amount of slack that can be pulled out. In this
case, one could argue that entropic stiffening would be the main
origin of stiffening. However, the DN calculations show that at these
densities stiffening lies in the network rather than in its
constituents: at small strains the mechanical response is dominated by
bending/buckling of filaments and nonaffine rotations and
translations. These local network rearrangements induce a transition
from a bending-dominated regime at small strains to a
stretching-dominated regime at larger strains in which strings of
connected filaments dominate the mechanical response. At higher
network densities,  the number of cross-links increases and the mean
cross-link distance $\lc$ decreases, so that the shorter segments
between cross-links contain less slack. The higher network density then
results in a mechanical response that is more affine in nature, with
the mentioned reduction in slack increasing the enthalpic stretching
contribution of the segments. By accounting for the reduced slack in
the AN calculations, the DN results were severely overestimated,
showing that, for the densities investigated, the discrete-network
architecture plays a key role. Only in the limit of extremely high
densities do the DN and reduced-slack AN calculations coincide. 

Hence, we conclude that, for physiological relevant densities,
stiffening in 2D networks results from nonlinearities in the network
response, rather than from entropic stiffening of individual filaments. Note that these conclusions might be different for three-dimensional networks, since the amount of slack in filaments is significantly higher than in 2D. Current work focuses on this issue in 3D networks.

\section*{Acknowledgments}

\noindent
The authors would like to acknowledge C. Storm (University of Leiden), F.C. MacKintosh 
(Vrije Universiteit, Amsterdam) and D. Weitz (Harvard University) for many fruitful discussions.

\section*{Appendix A}

\noindent
While real semiflexible filaments are, evidently, three-dimensional
objects that are often idealized as rods of diameter $d$, 2D models
require some 'projected' thickness $t$ for the specification of the
axial and the flexural stiffness in terms of
Young's modulus $E$, the area of a filament's cross section, $A$, and its second moment of
inertia, $I$. This 'projection', however, is
not trivial nor unique. While the moment of inertia of a circular
cross-section of diameter $d$ is $I=\pi d^4/64$, that of a projected
filament with thickness $t$ is $I=bt^3/12$ with some out-of-plane
width $b$. Aiming at actin filaments which typically have a diameter
of $d=8$ nm, we use 2D filaments with
$t=8$~nm. When a typical value of $E=2$~GPa is used for the
Young's modulus of actin, the bending stiffness $\kappaf$ 
of such 2D filaments
with a unit out-of-plane thickness ({i.e.}, $b=1$~m) is
\begin{equation}
\kappaf=EI=E\frac{bt^3}{12}\approx8.53\times10^{-17}~\mbox{Nm}^2.
\label{eq:kappaf}
\end{equation}
If we would use this result in Eq.~(\ref{lp2}), the persistence length
would be $2.1\times10^4$~m, much larger than the contour length
$L_{\rm C}$ and filaments would simply be straight, which they are not
in 3D. Therefore, the persistence length is scaled back to the contour length 
(scale factor: $4.8\times10^{-10}$) to generate semiflexible filaments for which
$L_{\rm P}=L_{\rm C}=10~\mu$m, as described in section II.

For the calculations performed in sections II, III, and IV we use the same
input value of $L_{\rm P}=10~\mu$m to generate the undulated filaments
({i.e.}, the initial configurational state \{$a_n^0$\},
determined by this value of $L_{\rm P}$) , but for the bending
stiffness $\kappaf$ in the beam-equation [Eq.~(\ref{beamequation})] we use the value
calculated in Eq.~(\ref{eq:kappaf}). For this reason $L_{\rm P}$ and $\kappaf$ are 
treated as separate parameters in the eqs.~(\ref{meansquaredan1})--(\ref{fcdepetedAVG4}).  
With the same 2D representation as underlying Eq.~(\ref{eq:kappaf}), the value of the 
stretching stiffness $EA$ is
\begin{equation}
\mu=EA=Ebt\approx16~\mbox{N}.
\label{eq:ea}
\end{equation}

\section*{Appendix B}

\noindent
For a Gaussian distribution with mean $\left<a_n^0\right>=0$ and standard deviation $s_n$ , 
the probability $p(a_n^0)\d a_n^0$ of finding a mode amplitude between $a_n^0$ and $a_n^0+
\d a_n^0$ is  
\begin{displaymath}
p(a_n^0)\d a_n^0=\frac{1}{\sqrt{2 \pi s_n^2}}\exp\left[-\frac{\left(a_n^0\right)^2}{2s_n^2}
\right]\d a_n^0.
\end{displaymath}
The mean-squared value of $a_n^0$ is given by
\begin{displaymath}
\left<\left(a_n^0\right)^2\right> = \int_{-\infty}^{\infty}\left(a_n^0\right)^2 p(a_n^0)\d 
a_n^0=s_n^2.
\end{displaymath}
Using Eq.~(\ref{meansquaredan1}) resulting from equipartition of the internal bending energy, 
the standard deviation is given by
\begin{displaymath}
s_{n}=\sqrt{\left<\left(a_n^0\right)^2\right>}=\sqrt{\frac{1}{L_{\rm P}}}q_n^{-1}~~~
\mbox{(for $n \ge 1$)}.
\end{displaymath}

\section*{Appendix C}

\setcounter{equation}{0}
\renewcommand{\theequation}{C-\arabic{equation}}
\noindent
In the absence of an externally applied force, a 2D semiflexible filament characterized by 
the tangent angle $\theta$
\begin{equation}
\theta(s,\{a_n^0\})=\sqrt{\frac{2}{L_{\rm C}}}\sum_{n\ge1}a_n^0\cos\left[q_ns\right],
\label{appc1}
\end{equation}
is in a configurational state denoted by \{$a_n^0$\}. The probability of a filament being in 
a state between \{$a_n^0$\} and d\{$a_n^0$\} is
\begin{equation}
p(\{a_n^0\})\textrm{d}\{a_n^0\}=\frac{1}{Z}\exp\left[-\frac{\mathscr{H}_B(\{a_n^0\})}{k_B T}
\right]\prod_{n \ge 1}\textrm{d}a_n^0,
\label{appc2}
\end{equation}
where $\textrm{d}\{a_n^0\}\equiv\prod_{n \ge 1}\textrm{d}a_n^0$, $Z$ is the partition 
function given by 
\begin{equation}
Z=\int_{-\infty}^{\infty}\ldots\int_{-\infty}^{\infty}\exp\left[-\frac{\mathscr{H}_B(\{a_n^0\})}
{k_B T}\right]\prod_{n \ge 1}\textrm{d}a_n^0,
\label{appc3}
\end{equation}
and $\mathscr{H}_B$ is the internal bending energy of the filament given by Eq.~(\ref{hamiltonian}). 
After substituting the Fourier series for $\theta$ into $\mathscr{H}_B$ and by using $\sin \theta 
\approx \theta$, one finds
\begin{equation}
\frac{\mathscr{H}_B(\{a_n^0\})}{\kB T}=\frac{L_{\rm P}}{2}\sum_{n \ge 1}q_n^2 (a_n^0)^2.
\label{appc4}
\end{equation}
This result can be substituted into Eq.~(\ref{appc3}) yielding the partition function
\begin{equation}
Z=\prod_{n \ge 1}~\int\limits_{-\infty}^{\infty}\exp\left[-\frac{L_{\rm P}}{2}q_n^2(a_n^0)^2\right]
\textrm{d}a_n^0=\prod_{n \ge 1}\left(\frac{2\pi}{L_{\rm P}q_n^2}\right)^{1/2}.
\label{appc5}
\end{equation}
The ensemble-average of a quantity $A(\{a_n^0)\})$ is defined by
\begin{equation}
\left<A\right>=\int\limits_{-\infty}^{\infty}\ldots\int\limits_{-\infty}^{\infty}p(\{a_n^0)\})A(\{a_n^0)\})
\prod_{n \ge 1}\textrm{d}a_n^0
\label{appc6}
\end{equation}
The end-to-end distribution function can be calculated by the following ensemble 
average\cite{WilhelmPRL96}
\begin{equation}
\mathscr{G}(r_0)=\left<\delta\left(r_0-\left[L_{\rm C}-\frac{1}{2}\sum_{n \ge 1}(a_n^0)^2\right]\right)\right>,
\label{appc7}
\end{equation}
where $\delta(r_0)$ is the Dirac delta function, and $L_{\rm C}-\frac{1}{2}\sum_{n \ge 1}(a_n^0)^2$ 
is the end-to-end distance of the chain in the absence of an externally applied force ($f_c=0$), 
according to Eq.~(\ref{fcdepeted}). Next, we can combine Eqs.~(\ref{appc2},\ref{appc4}-\ref{appc7}) 
and make use of the following property of the delta function
\begin{displaymath}
\delta(r_0)=\frac{1}{2\pi}\int\limits_{-\infty}^{\infty}\exp[izr_0]\textrm{d}z,
\end{displaymath}
resulting in the following integral solution for $\mathscr{G}(r_0)$:
\begin{equation}
\mathscr{G}(r_0)=\frac{1}{2\pi}\int\limits_{-\infty}^{\infty}\textrm{d}z\exp\left[iz(r_0-L_{\rm C})\right]
\prod_{n \ge 1}\left(\frac{L_{\rm P}q_n^2}{L_{\rm P}q_n^2-iz}\right)^{1/2}
\label{appc8}
\end{equation}
Next, we rewrite the  product inside the integral expression as
\begin{equation}
\prod_{n \ge 1}\left(\frac{L_{\rm P}q_n^2}{L_{\rm P}q_n^2-iz}\right)^{1/2}=
\exp\left[\frac{1}{2}\psi(iz)\right],
\label{appc9}
\end{equation}
in which $\psi$ is defined as
\begin{displaymath}
\psi(z)=\sum_{n \ge 1} \ln\left(\frac{L_{\rm P}q_n^2}{L_{\rm P}q_n^2-z}\right)
\end{displaymath}
The derivative $\d \psi/\d z$ can be written as
\begin{displaymath}
\frac{\d \psi(z)}{\d z}=\sum_{n \ge 1} \frac{1}{L_{\rm P}q_n^2-z}=\frac{1}{2z}\left[1-\sqrt{\frac{z L_{\rm C}^2}{L_{\rm P}}}\cot\sqrt{\frac{z L_{\rm C}^2}{L_{\rm P}}}\right],
\end{displaymath}
and integrated to yield (with $\psi(0)=0$)
\begin{equation}
\psi(z)=\ln\left(\frac{\sqrt{z L_{\rm C}^2/L_{\rm P}}}{\sin\sqrt{z L_{\rm C}^2/L_{\rm P}}}\right).
\label{appc10}
\end{equation}
Equations (\ref{appc9},\ref{appc10}) are subsituted into Eq.~(\ref{appc8}), and by substitution of variables accoring to $z=iL_{\rm P}\omega^2$ we eventually arrive to
\begin{eqnarray}
\mathscr{G}(r_0)=\frac{L_{\rm P}}{2\pi i}\int\limits_{-i\infty}^{i\infty}\textrm{d}\omega \exp\left[-\omega^2L_{\rm P}(r_0-L_{\rm C})\right] \nonumber\\
\times~2\omega\left(\frac{\omega L_{\rm C}}{\sinh(\omega L_{\rm C})}\right)^{1/2}
\label{appc11}
\end{eqnarray}
Next, we can use the binomial expansion\cite{binomial} to expand the inverse square root of $\sinh(\omega L_{\rm C})$
\begin{displaymath}
\sinh(\omega L_{\rm C})^{-\half}=\sqrt{2}\sum_{k\ge 0} {-\half \choose k}(-1)^k\exp\left[-(2k+\half)\omega L_{\rm C}\right].
\end{displaymath}
The distribution function then becomes
\begin{eqnarray}
\mathscr{G}(r_0)=\frac{L_{\rm P}\sqrt{2L_{\rm C}}}{\pi i}\sum_{k \ge 0} {-\frac{1}{2} \choose k}(-1)^k \int\limits_{-i\infty}^{i\infty}\textrm{d}\omega \omega^{\frac{3}{2}} \nonumber\\
\times~ \exp\left[-\omega^2 L_{\rm P}(r_0-L_{\rm C})-(2k+\frac{1}{2})\omega L_{\rm C}\right] 
\label{appc12}
\end{eqnarray}
We can rewrite Eq.~(\ref{appc12}) using so-called {parabolic cylinder
functions} $D_{\nu}(z)$, defined by\cite{Abramowitz}
\begin{equation}
D_{\nu}(x)=\frac{1}{\sqrt{2\pi}i}\exp\left[\frac{x^2}{4}\right]\int\limits_{-i\infty}^{i\infty}s^{\nu}\exp\left[-xs+\frac{1}{2}s^2\right]\textrm{d}s. 
\label{appc13}
\end{equation}
Using the following substitution of variables,
\begin{displaymath}
s=\omega\sqrt{2L_{\rm P}(L_{\rm C}-r_0)} \mbox{, and }
x=\frac{2(k+1/4)L_{\rm C}}{\sqrt{2L_{\rm P}(L_{\rm C}-r_0)}},
\end{displaymath}
the distribution function is rewritten as (with $\nu=3/2$)
\begin{eqnarray}
\mathscr{G}(r_0)=\frac{2L_{\rm P}\sqrt{L_{\rm
C}}}{\sqrt{\pi}}\sum_{k\ge0}{-\frac{1}{2} \choose k}(-1)^k
\frac{1}{[2L_{\rm P}(L_{\rm C}-r_0)]^{5/4}} \nonumber\\
\times~\exp\left[-\frac{(k+\frac{1}{4})^2L_{\rm C}^2}{2L_{\rm P}(L_{\rm C}-r_0)}\right]D_{3/2}
\left[\frac{2(k+\frac{1}{4})L_{\rm C}}{\sqrt{2L_{\rm P}(L_{\rm C}-r_0)}}\right] \nonumber\\
\label{appc14}
\end{eqnarray}
Finally, we can rewrite the binomial coefficient in terms of the {Gamma} function:
\begin{displaymath}
{-\frac{1}{2} \choose k}=\frac{\Gamma\left(\frac{1}{2}\right)}{\Gamma(k+1)\Gamma\left(\frac{1}{2}-k\right)},
\end{displaymath}
with $\Gamma(k+1)=k!$. By making use of the following identity of the Gamma function
\begin{displaymath}
\Gamma\left(n+\frac{1}{2}\right)=\frac{(2n-1)!!}{2^n}\Gamma\left(\frac{1}{2}\right),
\end{displaymath}
substituting $n=-k$ and the identity
\begin{displaymath}
(-2k-1)!!=\frac{(-1)^k}{(2k-1)!!},
\end{displaymath}
we arrive at the the following expression for the binomial coefficient
\begin{displaymath}
{-\frac{1}{2} \choose k}=\frac{(2k-1)!!}{2^kk!}(-1)^k.
\end{displaymath}
The distribution function is therefore written as
\begin{eqnarray}
\mathscr{G}(r_0)=\frac{2L_{\rm P}\sqrt{L_{\rm
C}}}{\sqrt{\pi}}\sum_{k\ge0}\frac{(2k-1)!!}{2^kk!}
\frac{1}{[2L_{\rm P}(L_{\rm C}-r_0)]^{5/4}} \nonumber\\
\times~\exp\left[-\frac{(k+\frac{1}{4})^2L_{\rm C}^2}{2L_{\rm P}(L_{\rm C}-r_0)}\right]D_{3/2}
\left[\frac{2(k+\frac{1}{4})L_{\rm C}}{\sqrt{2L_{\rm P}(L_{\rm C}-r_0)}}\right] \nonumber\\
\end{eqnarray}

\section*{Appendix D}

\setcounter{equation}{0}
\renewcommand{\theequation}{D-\arabic{equation}}
\noindent
We consider a `network' consisting of parallel filament strings that are initially normal to the
plane of shear, separated by a distance $H$ [see dashed lines in sketch(a)]. The strings have a stretching 
stiffness $\mu$ and connect the top and bottom plates separated by a distance $W$. Next, the network is subjected 
to a shear of strain $\Gamma$ by displacing the top plate by a distance $u$. Strings rotate by an angle 
$\beta=\arctan\left(\Gamma\right)$ and undergo a stretch, leading to a certain shear stress $\tau$. 
\begin{figure}[h] % sketch
\begin{center}
\includegraphics[width=0.45\textwidth]{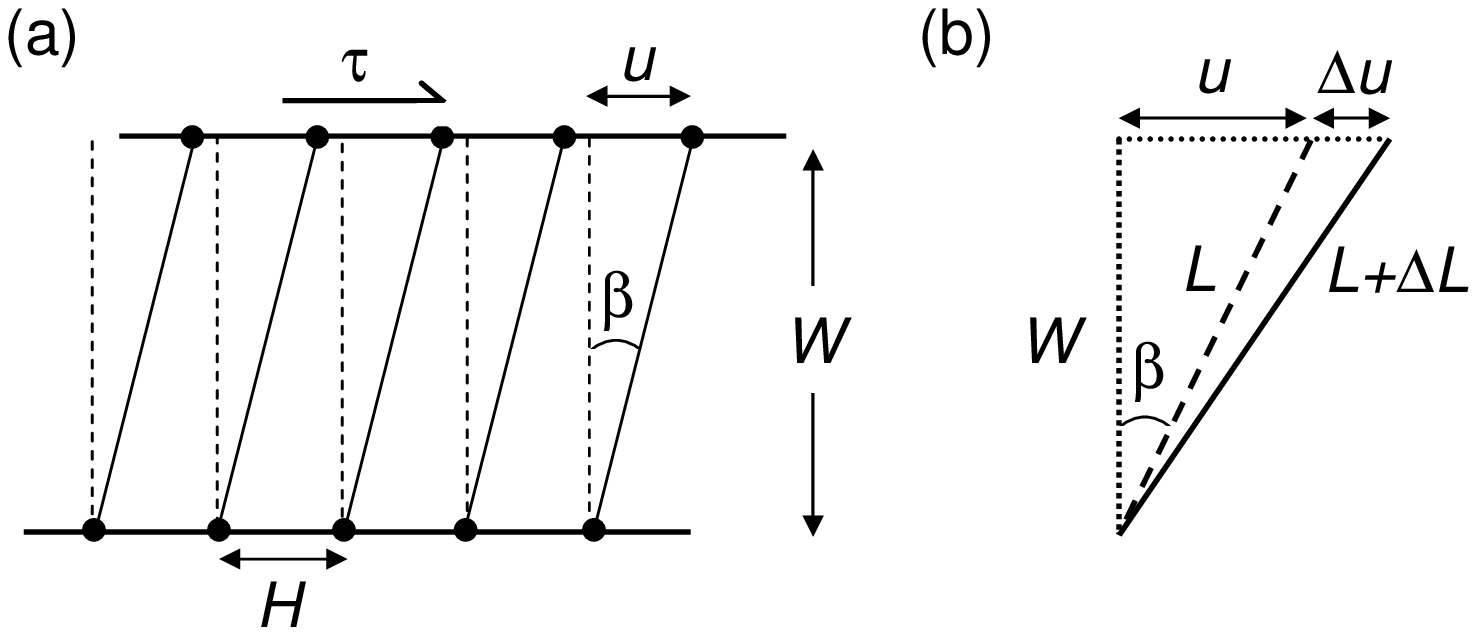}
\end{center}
\end{figure}
The instantenous stiffness $G$ can be found by increasing the top-plate displacement from $u \mbox{ to } u+\Delta u$ [see sketch(b)]
corresponding to a increase $\Delta \Gamma$ in strain and resulting in an increase $\Delta \tau$ in stress. Using
a unit out-of-plane dimension, the shear stiffness is defined as
\begin{equation}
G=\frac{\Delta \tau}{\Delta \Gamma}=\frac{W}{H}\frac{\Delta P_x}{\Delta u},
\label{appd1}
\end{equation} 
in which $\Delta P_x$ is the increase in string's force ($\Delta P$) projected on the shearing direction, i.e., 
$\Delta P_x=\Delta P \sin\beta$. The linear response of each string,
\begin{equation}
\Delta P=\mu\frac{\Delta L}{L},
\label{appd2}
\end{equation} 
in terms of its elongation $\Delta L$, along with the geometrical
relations $\Delta L=\Delta u \sin\beta$ and $L=u/\sin\beta$, yields
\begin{equation}
\Delta P_x=\mu\frac{\Delta u}{u}\sin^3\beta.
\label{appd3}
\end{equation} 
Inserting Eq.~(\ref{appd3}) into Eq.~(\ref{appd1}) and using $u=W\tan\beta$, we arrive at
\begin{equation}
G=\frac{\mu}{H}\sin^2\beta\cos\beta.
\end{equation} 
Finally, since the distance between strings is inversely proportional
to the network's line density, i.e., $H\propto\rho^{-1}$, and by
employing $\beta=\arctan(\Gamma)$ we obtain at the following scaling relation for the stiffness
\begin{equation}
G \propto \rho \mu \frac{\Gamma^2}{(1+\Gamma^2)^{3/2}} \,.
\end{equation}


\begin{thebibliography}{9}

\bibitem{Wang95} N. Wang and D.E. Ingber, Biochemistry and Cell
Biology, \textbf{73}, 327--335 (1995).

\bibitem{GardelScience04} M.L. Gardel, J.H. Shin, F.C. MacKintosh, L.
Mahadevan, P. Matsudaira, D.A. Weitz, Science \textbf{304}, 1301
(2004).

\bibitem{Xu00} J. Xu, Y. Tseng, D. Wirtz, J. Biol. Chem.
\textbf{275}, 35886 (2000).

\bibitem{Tseng04} Y. Tseng, K.M. An, O. Esue, D. Wirtz, J. Biol. Chem.
\textbf{279}, 1819 (2004).

\bibitem{Janmey91} P.A. Janmey, U. Euteneuer, P. Traub and M.
Schliwa, J. Cell Biol. \textbf{113}, 155 (1991).

\bibitem{Ma99} L. Ma, J. Xu, P.A. Coulombe  and D.J. Wirtz, J. Biol.
Chem. \textbf{274}, 19145 (1999).

\bibitem{Leterrier96} J.F. Leterrier, J. K\"{a}s, J. Hartwig, R.
Vegners, P.A. Janmey,  J. Biol. Chem. \textbf{271}, 15687 (1996).

\bibitem{Bale88} M.D. Bale, J.D. Ferry, Thromb. Res. \textbf{52},
565 (1988).

\bibitem{Janmey83} P.A. Janmey, E. Amis, J. Ferry, J. Rheol.
\textbf{27}, 135 (1983).

\bibitem{Wachsstock94} D.H. Wachsstock, W.H. Schwartz, and T.D. Pollard,
Biophys. J. \textbf{66}, 801 (1994).

\bibitem{WagnerPNAS06} B. Wagner, R. Tharmann, I. Haase, M. Fischer, and A.R. Bausch,
Proc. Natl. Acad. Sci. USA \textbf{103}, 13974 (2006).

\bibitem{GardelPNAS06} M.L. Gardel, F. Nakamura, J.H. Hartwig, J.C. Crocker, T.P. Stossel, and D.A. Weitz,
Proc. Natl. Acad. Sci. USA \textbf{103}, 1762 (2006).

\bibitem{DiDonnaPRL06} B.A. DiDonna and A.J. Levine,
Phys. Rev. Lett. \textbf{97}, 068104 (2006).

\bibitem{McGrath06} J.L. McGrath, Current Biology \textbf{16}, R326 (2006).

\bibitem{Tharmann} R. Tharmann, M.M.A.E Claessens, and A.R. Bausch, submitted.

\bibitem{MacKintoshPRL95} F.C. MacKintosh, J. K\"{a}s and P.A. Janmey,
Phys. Rev. Lett. \textbf{75}, 4425 (1995).

\bibitem{HeadPRL03} D.A. Head, A.J. Levine and F.C. MacKintosh, Phys. Rev. Lett. \textbf{91}, 
108102 (2003).

\bibitem{HeadPRE03} D.A. Head, A.J. Levine and F.C. MacKintosh, Phys. Rev. E \textbf{68}, 
061907 (2003).

\bibitem{WilhelmPRL03} J. Wilhelm and E. Frey, Phys. Rev. Lett. \textbf{91},
108103 (2003).

\bibitem{Levine} A.J. Levine, D.A. Head, and F.C. MacKintosh, J. Phys.: Condens Matter \textbf{16},
s2079 (2004).

\bibitem{DiDonnaPRE05} B.A. DiDonna and T.C. Lubensky, Phys. Rev. E \textbf{72},
066619 (2005).

\bibitem{Storm} C. Storm, J.J. Pastore, F.C. MacKintosh, T.C. Lubensky, and P.A. Janmey, Nature
\textbf{435}, 191 (2005).

\bibitem{HeussingerPRL06} C. Heussinger and E. Frey, Phys. Rev. Lett. \textbf{97},
105501 (2006).

\bibitem{KroyPRL96} K. Kroy and E. Frey, Phys. Rev. Lett. \textbf{77}, 306 (1996).

\bibitem{Odijk} T. Odijk, Macromolecules \textbf{16}, 1340 (1983).

\bibitem{Liu} X. Liu and G.H. Pollack, Biophys. J. \textbf{83}, 2705 (2002).

\bibitem{Wu93} P.D. Wu and E. van der Giessen, J. Mech. Phys. Solids \textbf{41}, 427 (1993).

\bibitem{OnckPRL05} P.R. Onck, T. Koeman, T. van Dillen, and E. van der Giessen, Phys. Rev. Lett. \textbf{95}, 178102 (2005).

\bibitem{Howard} Howard, J., \textit{Mechanics of motor proteins and the 
cytoskeleton} (Sinauer Associates, Inc., Sunderland, Massachusetts, 2001).

\bibitem{Gittes93} F. Gittes, B. Mickey, J. Nettleton, and J. Howard, J. Cell Biol. \textbf{120}, 923 (1993).

\bibitem{rmrkNOFT} In principle $N \to \infty$, which we will use in
this paper's anlytical calculations. We will show, however, that the
shape of a filament can be described well by $N$-values on the order
of ten.

\bibitem{remarkendpoint} As can be seen from Fig.~\ref{figrndfil} the filament does not exactly end on the $x$-axis due to the approximation made in Eq.~(\ref{a0approx}). Therefore, the (small) value of $y(L_{\rm C})$ is also taken into account in calculating the end-to end distance. 

\bibitem{WilhelmPRL96}J. Wilhelm and E. Frey, Phys. Rev. Lett. \textbf{77}, 2581 (1996).

\bibitem{rmrkstatic} The adjective `static' here signifies that only the initial, undulated configuration is taken into account (no undulation dynamics).

\bibitem{rmrkfinalstiffness} This follows directly from Eq.~(\ref{fcdepeted3}): close to full stretching $r$ approaches $L_{\rm C}$, resulting in $r_{\mu}\rightarrow L_{\rm C}(1+\fc/\mu)$ and hence a stiffness of $\mu/L_{\rm C}$.

\bibitem{rmrkexpansion}  Based on Eq.~(\ref{etedist}), we can use the following approximation for semiflexible filaments: $\left<r_0\right> \approx \left<r_{0}^{2}\right>^{1/2}=2L_{\rm P}\sqrt{\left(2\left\{\exp\left[
-\alpha /2\right]-1\right\}+\alpha\right)}$ with $\alpha \equiv L_{\rm C}/L_{\rm P} \lesssim 1$.  Taylor expansion in $\alpha$ yields $\left<r_0\right> \approx L_{\rm P}\alpha -L_{\rm P}\alpha^2/12$. The average slack is therefore $L_{\rm C}^2/(12L_{\rm P})$.

\bibitem{Marko95} J.F. Marko, E.D. Siggia, Macromol. \textbf{28}, 8759 (1995).

\bibitem{rmrktilde}Here and in the sequel, the superposed tilde denotes quantities in the dynamic description.

\bibitem{Wu95} P.D. Wu and E. van der Giessen, Phil. Mag. A \textbf{71}, 1191 (1995).

\bibitem{rmrkvolconservation} An initial volume element $\d V$ subjected to a deformation process characterized by the strain gradient tensor ${\bf F}$, changes its volume to $\d V'=\det{\bf F}~\d V$. Volume conservation is thus expressed by $\det{\bf F}=1$.

\bibitem{rmrkcodf1} This is similar to the general expression of the chain-orientation-distribution-function (CODF) in three dimensions (with spherical angles $\theta$ and $\phi$) which reads $C(\theta,\phi;{\bf F})=C_0\lambda^3(\theta,\phi;{\bf F})$ with $C_0=1/(4\pi)$, as derived by Wu and Van der Giessen.~\cite{Wu93,Wu95}

\bibitem{rmrkcodf2} Note that the CODF is a periodic function of period $\pi$. Therefore the normalization in Eq.~(\ref{codf}) is done over an angular range of $\pi$, e.g. $[-\pi/2, \pi/2]$ or $[0, \pi]$, but the CODF is plotted here over the whole angular range of $2\pi$.

\bibitem{rmrkproduct} This limiting value can be obtained by considering $N_{\rm f}$ straight filaments of length $L_{\rm C}$, oriented at random angles $\alpha$ with the x-axis, in a square cell of dimension $W$. The probability of a filament intersecting the x-axis is simply $(L_{\rm C}/W) \sin \alpha$. The probability of this filament intersecting a filament aligned with the x-axis is thus $(L_{\rm C}/W)^2 \sin \alpha$. The average cross-link probability can be found by averaging over the angle $\alpha$ and results in $2 L_{\rm C}^2/(\pi W^2)$. In the high-density limit, the mean number of cross-links per filament, $L_{\rm C}/\lc$, is simply the product of the average cross-link probability and number of filaments, {i.e.}, $2 \Nf L_{\rm C}^2/(\pi W^2)$. Using $\rho=\Nf L_{\rm C}/W^2$, this results in the high-density limit $\rho \lc=\pi/2$.\cite{HeadPRE03rc}

\bibitem{HeadPRE03rc} D.A. Head, F.C. MacKintosh, and A.J. Levine, Phys. Rev. E \textbf{68}, 025101(R) (2003).

\bibitem{rmrkprefactor} Note that the prefactor in 
Eq.~(\ref{macroshearstressnorm}) changes to $\overline{\rho}\left<r_0\right>[\pi \mu\lc(\rho)]^{-1}$, with $\left<r_0\right>$ the average end-to-end distance of a segment. However, the dimensionless quantities
$\overline{\rho}$ and $\overline{\tau}$ are still defined as $\rho L_{\rm C}$ and $\sigma_{12} b L_{\rm C}/\mu$, respectively.

\bibitem{expansion} First-order Taylor expansion of the integrand in Eq.~(\ref{macroshearstressnorm2}) in terms of $\Gamma$: $\lambda^3(\phi;\Gamma)(\lambda(\phi;\Gamma)-1)\sin\phi\cos\phi \approx \Gamma \sin^2\phi\cos^2\phi$, resulting in 
$\overline{\tau}=(\overline{\rho}/8)\Gamma$.

\bibitem{binomial} $(1+x)^\alpha=\sum_{k\ge0}{a \choose k}x^k$ which holds for all complex $\alpha$ with $|\alpha|<1$.

\bibitem{Abramowitz} M. Abramowitz and I.A. Stegun, \textit{Handbook
of mathematical functions} (Dover Publ., New York, 1965).

\end{thebibliography}
\end{document}